\documentclass{aa}  
\usepackage{amsmath,amssymb}
\usepackage[breaklinks=true]{hyperref}
\usepackage{threeparttable}
\usepackage{graphicx}
\usepackage{xcolor}
\usepackage{listings}
\usepackage{txfonts}
\usepackage{tikz}
\usepackage{orcidlink}
\author{Wenbo Wu\,\orcidlink{0000-0002-3354-9492}\inst{\ref{NAOC},\ref{IAC},\ref{ULL}}, Xianhao Ye\,\orcidlink{0000-0002-5805-8112}\inst{\ref{NAOC},\ref{IAC},\ref{ULL}}, C. Allende Prieto\,\orcidlink{0000-0002-0084-572X
}\thanks{Corresponding author, e-mail: carlos.allende.prieto@iac.es}\inst{\ref{IAC},\ref{ULL}}, Yuqin Chen\,\orcidlink{0000-0002-8442-901X}\inst{\ref{NAOC},\ref{CAS}}, Xiang-Xiang Xue\,\orcidlink{0000-0002-0642-5689}\inst{\ref{NAOC},\ref{BNU}}, Gang Zhao\,\orcidlink{0000-0002-8980-945X}\thanks{Corresponding author, e-mail: gzhao@nao.cas.cn}\inst{\ref{NAOC},\ref{CAS}}, Jingkun Zhao\,\orcidlink{0000-0003-2868-8276}\inst{\ref{NAOC},\ref{CAS}}, David S. Aguado\,\orcidlink{0000-0001-5200-3973
}\inst{\ref{IAC},\ref{ULL}}, Jonay I. González Hernández\,\orcidlink{0000-0002-0264-7356}\inst{\ref{IAC},\ref{ULL}}, Rafael Rebolo\,\orcidlink{0000-0003-3767-7085}\inst{\ref{IAC},\ref{ULL}}}
\institute{National Astronomical Observatories, Chinese Academy of Sciences, Beijing 100101, People's Republic of China\label{NAOC}
\and School of Astronomy and Space Science, University of Chinese Academy of Sciences, Beijing 100049, People's Republic of China\label{CAS}
\and Instituto de Astrofísica de Canarias, C/ Vía Láctea s/n, E-38205 La Laguna, Tenerife, Spain \label{IAC}
\and Universidad de La Laguna, Departamento de Astrofísica, 38206 La Laguna, Tenerife, Spain \label{ULL}
\and Institute for Frontiers in Astronomy and Astrophysics, Beijing Normal University, Beijing 102206, People's Republic of China \label{BNU}}

\date{Received ? ?, ?; accepted ? ?, ?}
\abstract

\begin{document}

\title{Mapping the Milky Way with Gaia Bp/Rp spectra}
\subtitle{\uppercase\expandafter{\romannumeral2}: The inner stellar halo traced by a large sample of blue horizontal branch stars}
\titlerunning{Mapping the Milky Way with Gaia XP spectra \uppercase\expandafter{\romannumeral2}}
\authorrunning {Wu et al.}

\date{Received ? xx, xxxx}
  \abstract 
   {Due to their nearly constant absolute magnitudes and old ages, blue horizontal branch (BHB) stars are frequently used as standard candles to study the kinematics and structures of our galaxy. The number of identified BHB stars has significantly increased due to the advent of large scale surveys in the last two decades. Recently, Gaia DR3 was released including a catalog of around 220 million low-resolution spectra (Bp/Rp, or XP hereafter). These data have great potential for identifying many interesting stellar objects including BHB stars.}
   {We construct a full-sky BHB catalog from Gaia Bp/Rp spectra and use it to explore the shape of the inner stellar halo.}
   {We selected BHB stars based on synthetic photometry and stellar atmosphere parameters inferred from Gaia Bp/Rp spectra. We generated the synthetic SDSS broad-band $ugr$ and Pristine narrow-band CaHK magnitudes from Gaia Bp/Rp data. A photometric selection of BHB candidates was made in the $(u-g, g-r)$ and $(u-\mathrm{CaHK},g-r)$ color-color spaces. A spectroscopic selection in $T_\mathrm{eff}-\log g$ space was applied to remove stars with high surface gravity. The selection function of BHB stars was obtained by using the Gaia DR3 photometry. A non-parametric method that allows the variation in the vertical flattening $q$ with the Galactic radius, was adopted to explore the density shape of the stellar halo.}
   {We present a catalog of 44,552 high latitude ($|b|>20^\circ$) BHB candidates chosen with a well-characterized selection function. The stellar halo traced by these BHB stars is more flattened at smaller radii ($q=0.4$ at $r\sim8$ kpc), and becomes nearly spherical at larger radii ($q=0.8$ at $r\sim25$ kpc). Assuming a variable flattening and excluding several obvious outliers that might be related to the halo substructures or contaminants, we obtain a smooth and consistent relationship between $r$ and $q$, and the density profile is best fit with by a single power law with an index $\alpha=-4.65\pm0.04$.} 
   {}

   \maketitle
%
\section{Introduction}
Blue horizontal branch stars are old (age$\geq$10 Gyr), metal-poor ([M/H]$<-1$), and low-mass stars ($\sim0.5-1M_\odot$) that have evolved past the helium flash at the end of the red giant phase. BHB stars were first identified by a photometric study of the globular clusters (GCs) of M92 and M3 in the 1950s \citep{1952AJ.....57....4A}. These core-helium burning stars are photometrically A or B-types, and form a clear horizontal branch extending from the blue side of the instability strip to the location of the hot subdwarfs in the Hertzsprung–Russell (HR) diagram. Due to their nearly constant absolute magnitudes, BHB stars are frequently used as standard candles to trace distant Galactic structures and study the mass assembly history of the Milky Way: determining the distance and the age of the stellar populations in the Galactic halo and the GCs \citep[e.g.,][]{1991ApJ...375..121P,2011A&A...533A..59J,2013ApJ...775..134V,2016MNRAS.463.3169D,2017AJ....154..233S,2019ApJ...884...67W}, measuring the radial density profile and the velocity distribution of the stellar halo \citep[e.g.,][]{1998MNRAS.297..732S,2011MNRAS.416.2903D,2016MNRAS.463.3169D,2018MNRAS.481.5223T,2018PASJ...70...69F,2019PASJ...71...72F,2021ApJ...919...66B,2022MNRAS.510.4706L,2025PASJ...77..178F,2024A&A...690A.166A,2024ApJ...975...81Y}, studying the kinematic substructures and the stellar streams \citep[e.g.,][]{2011ApJ...738...79X,2011ApJ...731..119R,2016MNRAS.456..602B,2018ApJ...862L...1D,2019ApJ...880...65Y,2019ApJ...886..154Y,2021ApJ...910..102G}, and estimating the enclosed mass of the Milky Way \citep[e.g.,][]{2008ApJ...684.1143X,2012ApJ...761...98K,2021MNRAS.501.5964D,2022MNRAS.516..731B}.

Photometrically, BHB stars reside in a narrow color range that facilitates their identification in specific color-color diagrams. \cite{1982AJ.....87.1515P} published a small catalog of A and B-type stars selected from the ($U-B,B-V$) diagram. \cite{1989MNRAS.238..225S} constructed a sample of 126 faint A-type stars (mainly BHB stars) satisfying $0\leq(B-V)_0\leq0.2$ and $V\leq19$. \cite{1989MNRAS.238..225S} selected 4,175 candidates of field horizontal-branch and A-type stars using objective-prism/interference-filter data. However, BHB catalogs from these early studies suffer from heavy contamination of high surface gravity stars such as blue stragglers (BS) and A-type main sequence (MS) stars. In the last two decades, the study of BHB stars has been largely advanced by large-scale sky surveys. \cite{2000ApJ...540..825Y} demonstrated that a filter cut using the dereddened color-color diagram $(u_0-g_0,g_0-r_0)$ can separate the high surface gravity stars from the BHB stars to a certain extent in the Sloan Digital Sky Survey \citep[SDSS;][]{2000AJ....120.1579Y}. \cite{2010A&A...522A..88S} identified 27,074 probable BHB stars from SDSS Data Release 7 (DR7) photometric data through a support vector machine (SVM) method. \cite{2012AJ....143...86V} found that the SDSS $z$ band photometry can also be used as a surface gravity indicator to distinguish BHB stars from contaminants. \cite{2007ApJS..168..277B} provided the coordinates and photometric information of 12,056 stars identified in the HK Survey as field horizontal-branch or A-type stars. \cite{2019ApJ...872..206M} detect 12,544 probable BHB stars in the Milky Way bulge-halo transition region by applying specific cuts in the $ZYJHK_s$ near-infrared bands from the Vía Láctea (VVV) ESO Public Survey data \citep{2010NewA...15..433M}. \cite{2019MNRAS.490.5757S} find that a combination of the Pristine survey CaHK narrow-band photometry and the SDSS $ugr$ broad-band photometry can achieve a high completeness (91\%) and purity (93\%) in the selection of BHB stars. \cite{2021A&A...654A.107C} present a catalog of 57,377 BHB candidates selected from the color-magnitude diagram (CMD) of Gaia Early Data Release 3 \citep[EDR3;][]{2021A&A...649A...1G} with a purity of $\sim70\%$. Their follow-up study selects 22,335 BHB stars from the CMD of Gaia DR3 and shows a contamination level of less than 12\% by the synthetic spectral energy distribution (SED) \citep{2024A&A...685A.134C}. \cite{2024A&A...690A.166A} identify 95,466 possible BHB stars from the Legacy Survey photometry data \citep{2019AJ....157..168D} using a probabilistic methodology.                  

A photometric selection will inevitably involve some contamination due to uncertainties in the dust maps and the mixture of BHB and BS stars in color-color space. The most accurate way to construct a clean catalog of BHB stars is via spectral features. The Balmer-line profiles have proven to be an efficient indicator to distinguish between BHB and BS stars, since their depth ($f_m$) and broadening ($D_{0.2}$) can provide indirect information of $T_\mathrm{eff}$ and $\log g$ \citep{1983ApJS...53..791P}. \cite{2002MNRAS.337...87C} developed a scale-width-shape method that discriminates the BHB stars from the contaminants based on a fitting of the Balmer-lines with a Sérsic profile. Through an analysis of the Balmer-line profiles and a color cut, previous studies selected 4,985 BHB stars from SDSS DR8 data \citep{2011ApJ...738...79X} and 5,436 BHB stars from the Large Sky Area Multi-Object Fiber Spectroscopic Telescope (LAMOST) DR5 data \citep{2024ApJS..270...11J}. As demonstrated in \cite{2022ApJ...940...30B}, the BHB and MS/BS stars can be distinguished by applying a Gaussian Mixture model to their $\log g$ distributions.      

Gaia Data Release 3 \citep[DR3;][]{2023A&A...674A...1G} has released around 220 million low-resolution spectra obtained from the blue ($330\leq\lambda\leq680\,\mathrm{nm}$) and red ($640\leq\lambda\leq1050\,\mathrm{nm}$) Gaia slitless spectrophotometers \citep{2021A&A...652A..86C,2023A&A...674A...3M,2023A&A...674A...2D}. These spectra (hereafter referred to as `XP spectra') have a variable resolution ranging from 20 to 90 as a function of wavelength. Through spectral shape and synthetic photometry, previous studies have identified many interesting stellar objects such as white dwarfs \citep{2023A&A...679A.127G,2024A&A...682A...5V}, ultracool dwarfs \citep{2024MNRAS.527.1521C}, carbon-enhanced metal-poor stars \citep{2023MNRAS.521.2745S}, stars at the tip of the red giant branch \citep{2023ApJ...950...83L}, and magnetic chemically peculiar stars \citep{2022A&A...667L..10P} from XP spectra. 

In this study, we aim to construct a clean, full-sky catalog of BHB stars selected from XP spectra. This paper is organized as follows. In Section~\ref{sec:identification}, we introduce the photometric and spectroscopic criteria for selecting BHB stars from the XP spectra. In Section~\ref{subsec:distance} and ~\ref{subsec:sf}, we present a clean BHB catalog with the heliocentric distance and the selection function provided. We analyze the density shape of the inner stellar halo traced by the obtained BHB catalog in Section~\ref{subsec:density}. Finally, a summary is made in Section~\ref{sec:summary}.    
\section{Construction of the BHB catalog}
\subsection{Identification}\label{sec:identification}
In this study, we will use synthetic photometry and stellar atmospheric parameters inferred from XP spectra to select probable BHB candidates. Gaia XP spectra are represented by a linear combination of Hermite basis functions \citep{2023A&A...674A...3M}. We used the Python library GaiaXPy\footnote[1]{\href{https://gaia-dpci.github.io/GaiaXPy-website/}{https://gaia-dpci.github.io/GaiaXPy-website/}}, developed by the Gaia Data Processing and Analysis Consortium (DPAC), to transfer these coefficients into calibrated photometry and spectra. 
\subsubsection{Photometric cuts}\label{subsec:photometric} GaiaXPy allows the generation of synthetic photometry for multiple photometric systems from the input internally calibrated XP spectra \citep{2023A&A...674A..33G}. Inspired by previous studies of \cite{2004AJ....127..899S} and \cite{2019MNRAS.490.5757S}, we used the synthetic SDSS and Pristine CaHK photometry to make a rough photometric selection of BHB stars. We chose the photometric system of SDSS\_Std in the generation of $ugr$ broad-band photometry because the procedure in GaiaXPy benefits from a standardization that minimizes zero-point differences and/or trends as a function of colors in this system. The extinction coefficients ($R_u=4.239,R_g=3.303,\mathrm{and}\,R_r=2.285$) are provided by \cite{2011ApJ...737..103S}, and the CaHK extinction $R_\mathrm{CaHK}=3.918$ is obtained from \cite{2024A&A...692A.115M}. The dust reddening E(B-V) is provided by the two-dimensional dust map dustmaps.sfd.SFDQuery \citep{1998ApJ...500..525S,2011ApJ...737..103S,2018JOSS....3..695M}.

We first applied a color cut $0.8<u_\mathrm{std;0}-g_\mathrm{std;0}<1.4$ and $-0.3<g_\mathrm{std;0}-r_\mathrm{std;0}<0.1$, and tagged stars satisfying this criterion as $ugr$-cut candidates. In this study we used the subscript ";0" or "0" to indicate dereddened stellar color or extinction corrected magnitude. This photometric selection effectively removes some undesired objects, such as white dwarfs, quasars, and other stars \citep{2004AJ....127..899S,2012AJ....143...86V}. However, there are still some contaminants remaining in these $ugr$-cut candidates, particularly BS stars. 

\begin{figure}
    \centering
    \includegraphics[width = 8.8cm]{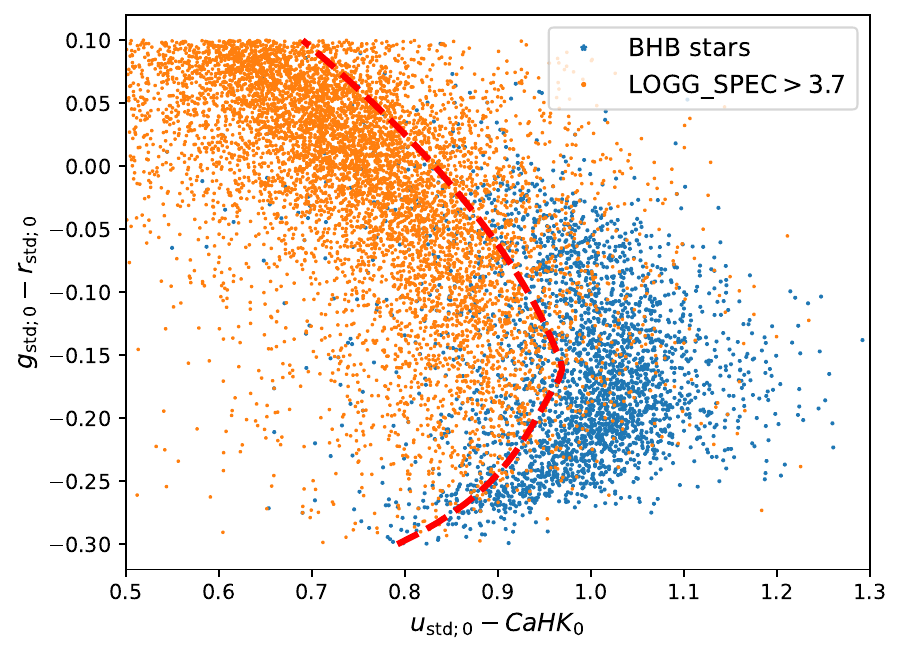}
    \caption{$u_\mathrm{std;0}-\mathrm{CaHK}_0$ versus $g_\mathrm{std;0}-r_\mathrm{std;0}$ for previously spectrally confirmed BHB stars (blue points) and high surface gravity stars (orange points). Although there are some mixtures of these two kinds of stars in this color-color diagram, we can still make a rough separation of them by the dashed red lines.}
    \label{fig:ucahk}
\end{figure}

The color-color space $(u_0-\mathrm{CaHK}_0, g_0-r_0)$ has been proven to efficiently separate BHB stars from BS stars by \citet{2019MNRAS.490.5757S}. However, our synthetic \textit{CaHK} photometry is not a standardized system like SDSS\_Std that may have some zero point offsets and trends as a function of color. Therefore, we need to carry out a test before applying this selection criterion with confidence. Among these $ugr$-cut candidates, 2,953 of them are previously confirmed as BHB stars by \cite{2022ApJ...940...30B}, and 7,781 of them are possible high surface gravity stars with $\mathrm{LOGG\_SPEC}>3.7$ in the Sloan Extension for Galactic Understanding and Exploration survey \citep[SEGUE;][]{2009AJ....137.4377Y}. Figure~\ref{fig:ucahk} shows the distribution of these BHB and high surface gravity stars in $(u_\mathrm{std;0}-\mathrm{CaHK}_0, g_\mathrm{std;0}-r_\mathrm{std;0})$ space. Although there is some overlap in this color-color diagram, we can see that these two kinds of stars are mainly located in two different regions. A dashed ridgeline $f(g_\mathrm{std;0}-r_\mathrm{std;0})$ was used to separate the BHB stars from the contaminants roughly. We applied this ridgeline to the whole $ugr$-cut sample and obtained 687,949 probable photometric-cut BHB candidates satisfying $0.8<u_\mathrm{std;0}-g_\mathrm{std;0}<1.4$, $-0.3<g_\mathrm{std;0}-r_\mathrm{std;0}<0.1$, and $u_\mathrm{std;0}-\mathrm{CaHK}_0 > f(g_\mathrm{std;0}-r_\mathrm{std;0})$, where the ridge line is approximately defined as:
\begin{equation}
\begin{aligned}
   f(g_\mathrm{std;0}-r_\mathrm{std;0})=&0.83-1.32(g_\mathrm{std;0}-r_\mathrm{std;0})-1.45(g_\mathrm{std;0}-r_\mathrm{std;0})^2\\&+11.24(g_\mathrm{std;0}-r_\mathrm{std;0})^3.
   \end{aligned}
\end{equation}

\subsubsection{Spectroscopic cuts}\label{subsec:spectrsoscopic}
Although photometric selection can remove a large number of undesired objects, there are still some remaining contaminants as illustrated in Figure~\ref{fig:ucahk}. To obtain a cleaner catalog, we fit the XP spectra of these photometric-cut BHB stars and inferred their stellar atmospheric parameters $(T_\mathrm{eff}, \log g, \mathrm{[M/H]})$.

Our sampled XP spectra have the same wavelength range of $360\,\mathrm{nm}\leq\lambda\leq990\,\mathrm{nm}$ as Paper-\uppercase\expandafter{\romannumeral1} of \citet{2025A&A...695A..75Y}, and contain 330 points that are spaced evenly on a logarithmic scale. To remove the effect of interstellar extinction, we corrected the XP data with the two-dimensional dust map dustmaps.sfd.SFDQuery and the extinction curve of ccm89 \citep{1989MNRAS.238..225S}. The FORTRAN-90 code FERRE\footnote[1]{\href{https://github.com/callendeprieto/ferre}{https://github.com/callendeprieto/ferre}} \citep{2006ApJ...636..804A}, which matches the input spectra to best-fitting templates from a grid of synthetic spectra by $\chi^2$ minimization, is used to infer stellar parameters from XP spectra ($T_\mathrm{eff},\log g, \mathrm{[M/H]}$). Since BHB stars are mainly in the temperature range of $7000<T_\mathrm{eff}<12000\,\mathrm{K}$, we adopted the nsc3 library of \cite{2018A&A...618A..25A} as model spectra in the fitting. To match the low resolution of the XP spectra, we degraded the synthetic spectra to a constant resolving power of $R=104$. Considering the lower temperature boundary of the nsc3 library, we need to ensure that the input XP spectra belong to stars of high temperature. In Paper-\uppercase\expandafter{\romannumeral1} we fit each whole XP spectrum with a library of spectra covering the temperature range of $3500<T_\mathrm{eff}<8000\,\mathrm{K}$. We required that $T_\mathrm{eff}$ estimated from Paper-\uppercase\expandafter{\romannumeral1} is larger than 6900 K.

We only kept stars with $G>11.5$ since \citet{2023A&A...674A..27A} mentioned that the internal-calibration process of XP spectra may have some instrumental effects for stars brighter than $G=11.5$. There is a high density of MS stars in the Magellanic clouds, potentially introducing significant contamination of MS stars to our selected BHB catalog. To correct for the effect of dust reddening in the XP spectra, we adopted $R_v=3.1$ in the extinction law of ccm89 for the Milky Way stars, which is not appropriate for stars in the Magellanic clouds. Therefore, following \cite{2021A&A...654A.107C}, objects were removed by excluding stars in the region of the Magellanic clouds where the local apparent stellar population density exceeds $50,000$ stars per $\deg^2$: (1) the Large Magellanic Cloud at $274.5^\circ<l<286.5^\circ$ and $-37.9^\circ<b<-27.9^\circ$. (2) the Small Magellanic Cloud at $299.8^\circ<l<305.8^\circ$ and $-46.3^\circ<b<-42.3^\circ$.
     
After fitting, we defined the residual between the Normalized XP spectra ($\mathrm{Flux_{XP}}$) and the fitting results ($\mathrm{Flux_{fitting}}$) as $\Delta \mathrm{Flux}$. Our sampled XP spectrum consists of 330 data points that are spaced evenly on a logarithmic scale within the wavelength range. To remove stars with poor fittings, we required that the number of points with $|\Delta \mathrm{Flux}|>0.1$ is less than 60 in any given spectrum.     

A metallicity cut $\mathrm{[M/H]}\leq-1$ is frequently used to remove metal-rich disk stars in a BHB sample. However, metallicity determined from such low-resolution spectra is possibly unreliable for stars with low ($T_\mathrm{eff}<4000$ K) and high ($T_\mathrm{eff}>7000$ K) temperatures. Therefore, we chose an alternative cut in galactic latitude $|b|>20^\circ$; 67,055 stars remain after the selection. 

Previous studies usually distinguish between BHB and BS stars by the shape of their Balmer-lines \citep{2008ApJ...684.1143X,2011ApJ...738...79X,2024ApJS..270...11J}. However, systematic errors that depend on the stellar color and magnitude are found in Gaia XP spectra \citep{2023A&A...674A...3M,2024ApJS..271...13H}, which could directly distort the shape of spectral lines and introduce a large uncertainty in the fitting of Balmer-lines. Paper-\uppercase\expandafter{\romannumeral1} introduces a method to correct for the effects of systematic errors, but it is not valid for stars with high temperatures. Since many previous studies have already obtained a reliable estimation of the stellar parameters without the correction of systematic errors in the XP spectra, we will rely on our estimated parameters rather than the profile of Balmer-lines in the selection. 

\begin{figure}
    \centering
    \includegraphics[width = 8.8cm]{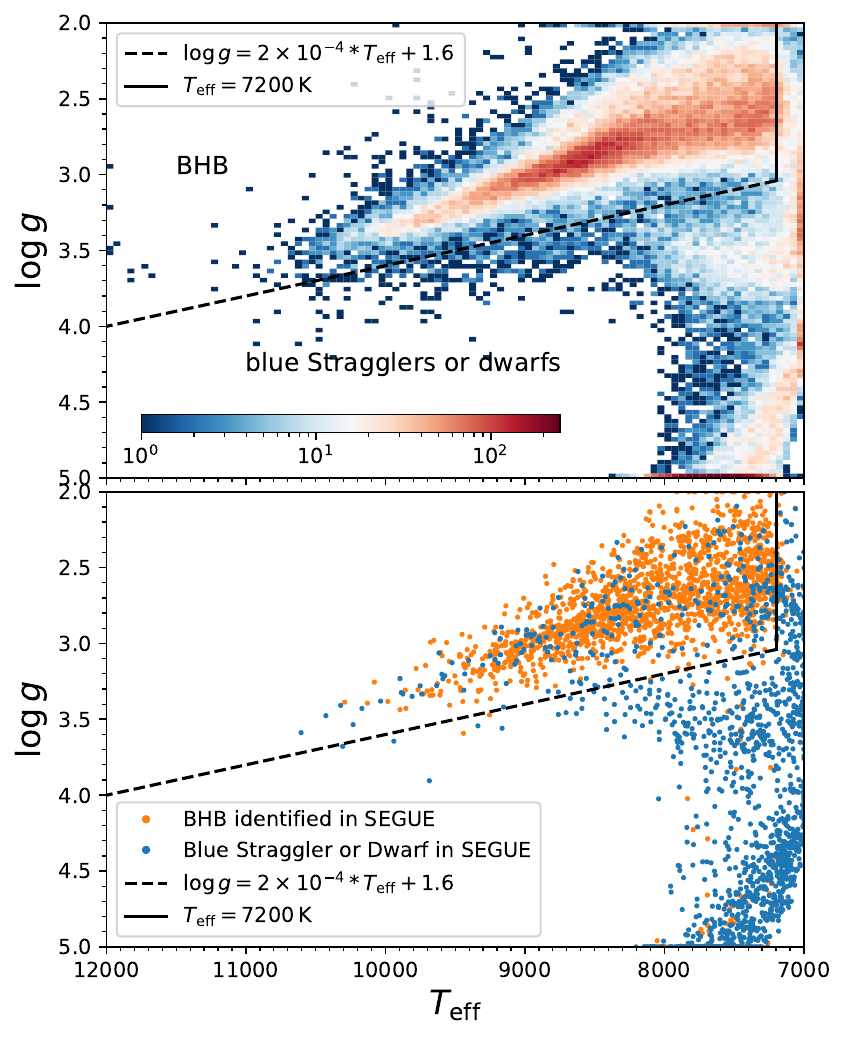}
    \caption{Separation of BHB stars from high $\log g$ contaminants in the $T_\mathrm{eff}-\log g$ space. The top panel shows the density distribution of the high-latitude photometric-cut BHB candidates in $T_\mathrm{eff}-\log g$ space. The bottom panel shows a small fraction of our candidates that have been previously identified as BHB (blue points) or high surface gravity stars (orange points) in the SEGUE data by \cite{2022ApJ...940...30B}. Two black lines are employed to separate these two types of stars, and we apply this cut to the whole candidates as shown in the top panel.}
    \label{fig:teff-logg}
\end{figure}

The top panel of Figure~\ref{fig:teff-logg} shows the density distribution of these high-latitude photometric-cut BHB candidates in $T_\mathrm{eff}-\log g$ diagram. We find that there are mainly three groups in Figure~\ref{fig:teff-logg}, of which the major group has the smallest $\log g$ and shows an increasing $T_\mathrm{eff}-\log g$ relationship, and the other two minor groups are composed of stars with higher $\log g$. According to the trends of $\log g$ with $T_\mathrm{eff}$, we suspect that stars in the major group are true BHB stars, while the other two minor groups are dwarfs or BS stars that are falsely identified as BHB stars in the photometric selection. To verify our assumption, we cross-matched our candidates with the SEGUE survey and found 3,292 common stars. The condition in the bottom panel of Figure~\ref{fig:teff-logg} is consistent with our assumption, as the majority of the previously spectrally confirmed BHB stars belong to the main group, while the other two groups are mainly composed of high surface gravity stars. A simple line of $\log g=2\times10^{-4}\times\,T_\mathrm{eff}+1.6$ was employed to separate the major group from the other two minor ones. We further removed stars with $T_\mathrm{eff}<7200\,\mathrm{K}$, as this selection line performs poorly at eliminating contaminants in the low-temperature regime. Although this separation is imperfect, as we can still see some overlap, it works well for most SEGUE stars and we applied it to all BHB candidates.    

The selection criteria in this section are summarized as Equation~\ref{eq:teff-logg}:
\begin{equation}
	\begin{aligned}
		&T_\mathrm{eff;Ye}\geq6900\,\mathrm{K}\\
	    &\mathrm{Stars\,in\,the\,region\,of\,the\,Magellanic\,Clouds\,are\,removed}\\
		&G>11.5\\
		&|b|>20^\circ\\
		&12000>T_\mathrm{eff}>7200\,\mathrm{K}\\
		&\log g<2\times10^{-4}\times\,T_\mathrm{eff}+1.6.
	\end{aligned}
	\label{eq:teff-logg}
\end{equation}
We defined these 44,552 selected stars as XP spectrally confirmed BHB candidates.

BHB stars projected within six half-mass radii from the center of known globular clusters within 25 kpc of the Sun were selected as potential globular cluster members. Parameters of the globular clusters were obtained from \cite{2021MNRAS.505.5957B} and \cite{2021MNRAS.505.5978V}. Our BHB catalog is composed of 44,032 field BHB stars and 520 potential globular cluster members.
          
\subsection{Distance estimation}\label{subsec:distance}
\begin{figure}
    \centering
    \includegraphics[width = 8.8cm]{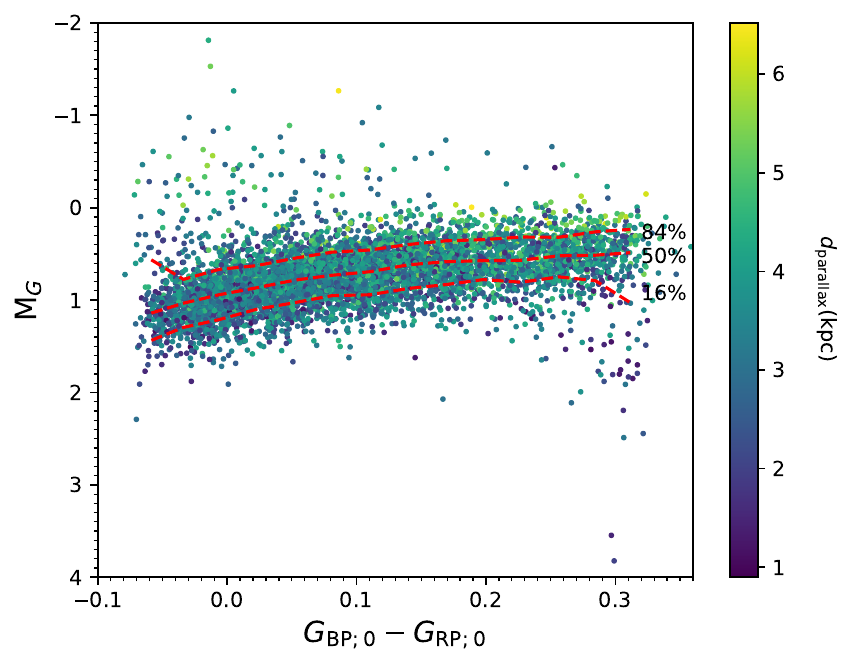}
    \caption{Gaia DR3 CMD of 6,065 selected BHB stars that have a reliable parallax. These BHB stars display a clear horizontal branch structure, and three dashed red curves that represent the $P_{16}, P_{50}, $ and $P_{84}$ percentiles distributions of $M_{G}$ are obtained.}
    \label{fig:cmd}
\end{figure}

Although BHB stars are known to have almost constant absolute magnitudes, there are still some variations as a function of color. \cite{2011MNRAS.416.2903D} fit the absolute magnitude $M_g$ of BHB stars as a function of $g-r$ color with a quartic polynomial. \cite{2021ApJ...910..102G} extent the fitting to the blue tail of the Horizontal Branch stars of $g-r<-0.25$. Due to the potential uncertainties in the synthetic $ugr$ photometry, we will rely on the Gaia broad band photometry to establish a relationship between the absolute magnitude $M_{G}$ and dereddened stellar color $G_\mathrm{BP;0}-G_\mathrm{RP;0}$. The extinction coefficients of Gaia Broad band photometry ($R_G=2.364, R_{G_\mathrm{BP}}=2.998, R_{G_\mathrm{RP}}=1.737$) are provided by \cite{2023ApJS..264...14Z}. The absolute magnitude $M_{G}$ was obtained from $M_{G}=G_0-5\log_{10}(d_\mathrm{parallax})+5$, where distance $d_\mathrm{parallax}$ is given in units of parsecs (pc) and derived from the Gaia DR3 parallax for 6,065 BHB stars with a reliable parallax (parallax\_over\_error>10). We took into account the parallax zero point offset of $-0.017$ mas as discussed by \cite{2021A&A...649A...4L}. These BHB stars exhibit a clear horizontal branch structure in the ($G_\mathrm{BP;0}-G_\mathrm{RP;0},M_{G}$) diagram as is shown in Figure~\ref{fig:cmd}. Three dashed red curves that represent the $P_{16}, P_{50}, $ and $P_{84}$ percentiles distributions were obtained after the removal of possible contaminants. We applied a quartic polynomial fitting to the three curves as Equation~\ref{eq:fitdistance}:  
\begin{equation}
\begin{aligned}
M_{G}(P_{84})=&0.68-1.11(G_\mathrm{BP;0}-G_\mathrm{RP;0})-26.49(G_\mathrm{BP;0}-G_\mathrm{RP;0})^2\\
&+182.44(G_\mathrm{BP;0}-G_\mathrm{RP;0})^3-326.44(G_\mathrm{BP;0}-G_\mathrm{RP;0})^4\\
M_{G}(P_{50})=&0.93-3.12(G_\mathrm{BP;0}-G_\mathrm{RP;0})+6.61(G_\mathrm{BP;0}-G_\mathrm{RP;0})^2\\
&+9.74(G_\mathrm{BP;0}-G_\mathrm{RP;0})^3-45.67(G_\mathrm{BP;0}-G_\mathrm{RP;0})^4\\
M_{G}(P_{16})=&1.17-3.28(G_\mathrm{BP;0}-G_\mathrm{RP;0})+16.95(G_\mathrm{BP;0}-G_\mathrm{RP;0})^2\\
&-86.45(G_\mathrm{BP;0}-G_\mathrm{RP;0})^3+181.85(G_\mathrm{BP;0}-G_\mathrm{RP;0})^4.
	\label{eq:fitdistance}
 \end{aligned}
\end{equation} 
The heliocentric distance, $d_\mathrm{helio}$, was calculated by substituting $M_{G}(P_{50})$ into $d_\mathrm{helio}=10^{0.2(G_0-M_{G}+5)}$.

\begin{figure}
    \centering
    \includegraphics[width = 8.8cm]{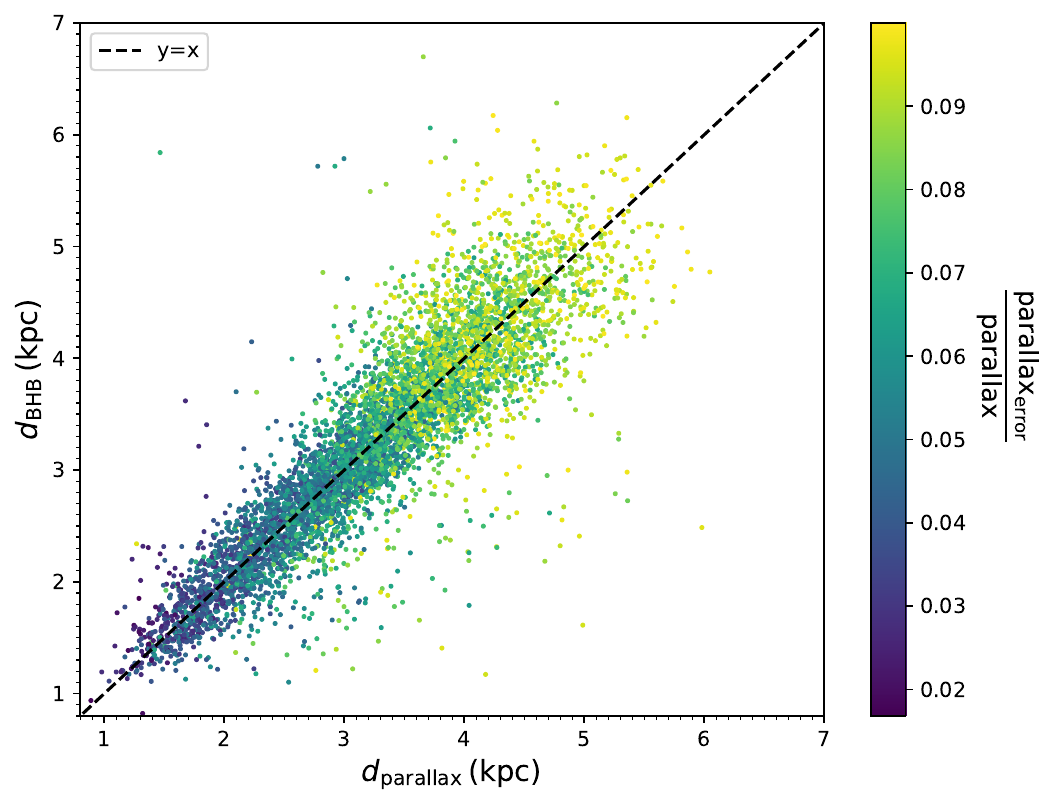}
    \caption{Comparison of the distance calculated from the parallax ($d_\mathrm{parallax}$) and the theoretical absolute magnitude ($d_\mathrm{helio}$) for BHB stars that have a reliable measurement of the parallax. An increasing dispersion with the increasing parallax error is expected.}
    \label{fig:comparedistance}
\end{figure}

\begin{figure}
    \centering
    \includegraphics[width = 8.8cm]{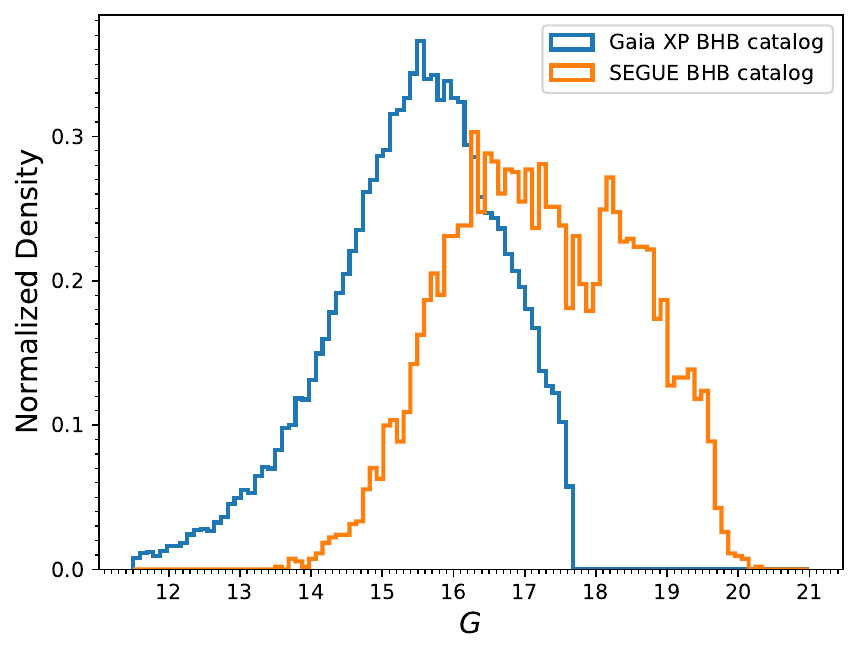}
    \caption{Normalized apparent magnitude distribution in $G$ band for XP spectrally confirmed BHB sample (blue hist) and SEGUE spectrally confirmed BHB sample of \citet{2022ApJ...940...30B} (orange hist). Our selected BHB sample is more focused on the inner stellar halo than the SEGUE BHB sample due to the limiting magnitude of Gaia XP spectra.}
    \label{fig:Gdensity}
\end{figure}

As is shown in Figure~\ref{fig:comparedistance}, our obtained $d_\mathrm{helio}$ is generally in line with $d_\mathrm{parallax}$. Figure~\ref{fig:Gdensity} shows the normalized apparent magnitude distribution of our selected BHB sample, with G magnitudes ranging from 11.5 to approximately 17.5, peaking at around 15.5. Due to the limiting magnitude of the XP spectra ($G\sim17.65$), $d_\mathrm{helio}$ of these BHB candidates is within 25 kpc, indicating that most of them are located in the inner halo (at a Galactocentric radius $r_\mathrm{gc}<30$ kpc).

\subsection{Completeness and purity}\label{subsec:sf}
To check the completeness and purity of our sample, we used the SEGUE spectrally confirmed BHB catalog of \cite{2022ApJ...940...30B} as the reference library. Many previous studies selected BHB stars from the SEGUE survey using different methods. Here we adopted the BHB sample of \cite{2022ApJ...940...30B} because their selection is also mainly based on the $\log g$ distribution like us. \citet{2019MNRAS.490.5757S} propose two different ways to estimate the completeness of their photometrically confirmed BHB sample selected from the Pristine survey. One approach consists of a direct comparison to the SEGUE spectrally confirmed BHB catalog, while the other is based on a test of Monte Carlo simulated mock data sets. The completeness estimated from the two methods is in good agreement for stars brighter than $g=19$. Although the SEGUE spectrally confirmed BHB catalog may not be perfectly complete, this consistency suggests that its completeness, $C(\mathrm{SEGUE})$), is likely close to 1 and does not vary significantly with the magnitude for stars of $g<19$. The limiting magnitude of Gaia XP spectra is around $G=17.65$, which is much brighter than $g=19$ according to the empirical transformation equations derived by \citet{2018A&A...616A...4E}.

\begin{figure}
    \centering
    \includegraphics[width = 8.8cm]{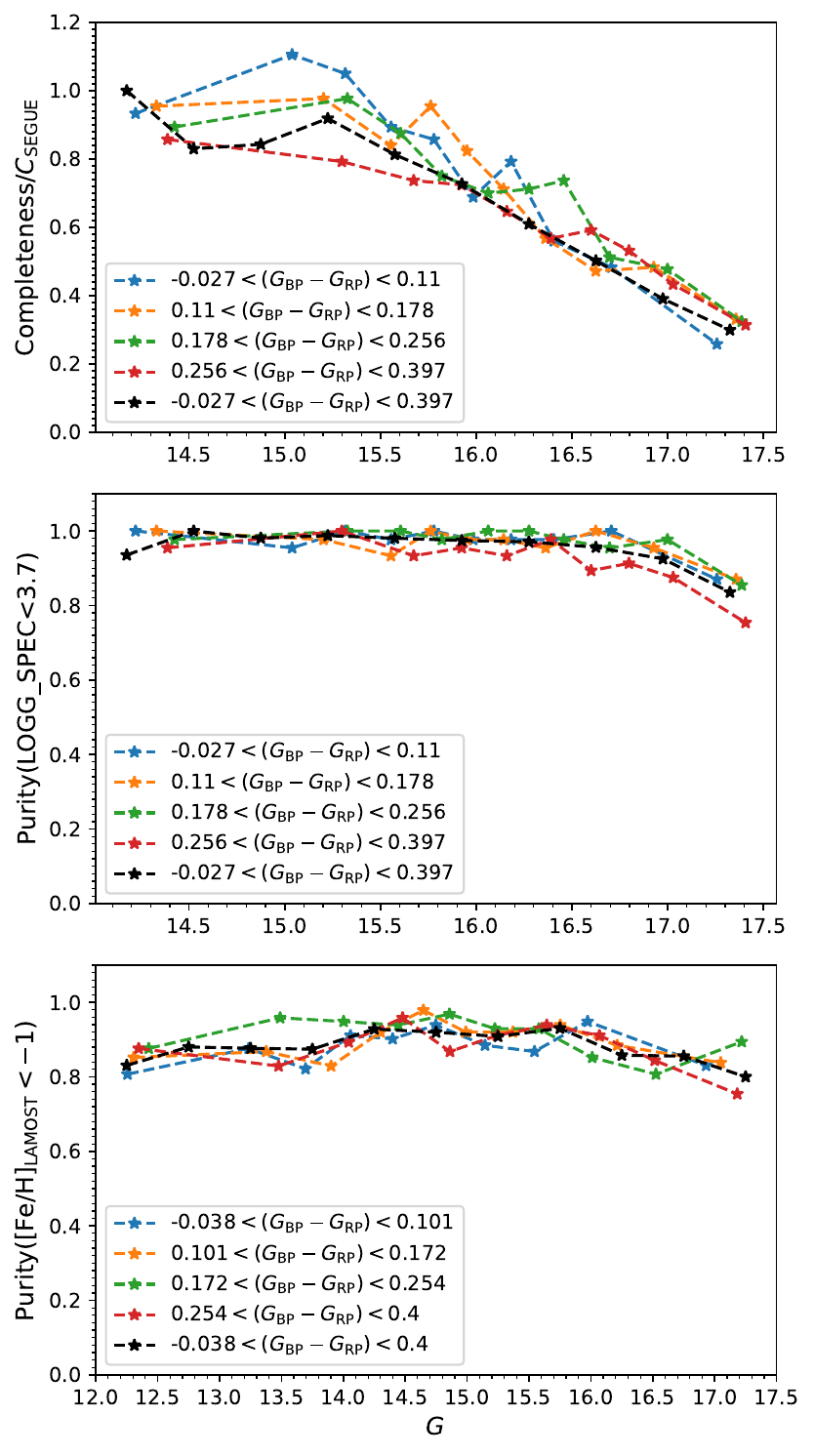}
    \caption{Completeness and purity of the overlapping BHB sample as a function of $G$ for BHB stars of different stellar colors. The top panel shows the ratio of BHB stars found from the same combined catalog (Gaia XP spectra$\times$SEGUE survey) by us and by \citet{2022ApJ...940...30B}. The middle panel presents the purity defined as the proportion of low $\log g$ stars in our 1,903 BHB candidates that have a measurement of LOGG\_SPEC in the SEGUE survey. The bottom panel displays the purity defined as the proportion of metal-poor stars in our 4,332 BHB candidates that have a measurement of $\mathrm{[Fe/H]_{LAMOST}}$ in the LAMOST OBA-type stars catalog. The dashed black lines describe the completeness and purity as a function of $G$ without the consideration of stellar color. The completeness declines rapidly as stars become fainter, while for the purity it remains almost constant with minimal variation.}
    \label{fig:compleAndPurity}
\end{figure}

To explore the completeness and purity of our sample, we constructed a combined catalog by cross-matching the Gaia XP spectra with the SEGUE survey. Within this catalog, 1,906 stars are identified as BHB candidates by Gaia XP spectra in this study. Among them, 1,815 candidates are confirmed to be real low $\log g$ stars with LOGG\_SPEC$<$3.7 according to the SEGUE data, which suggests an overall purity of around $1,815/1,906\sim95\%$ in this subsample. Cross-matching the SEGUE BHB sample of \citet{2022ApJ...940...30B} and the combined catalog yields 3,051 common sources. Among them, 1,643 are also identified as BHB stars by Gaia XP spectra in this study, while the rest are excluded by our photometric and spectroscopic cuts. We divided the 1,815 real low $\log g$ candidates into 40 bins of color and magnitude, and the bin edges were selected to ensure that each bin contains a similar number of around 45 stars. For each each bin, the top panel of Figure~\ref{fig:compleAndPurity} shows the ratio of BHB stars found from the same combined catalog by us (the 1,815 real low $\log g$ candidates) and by \citet{2022ApJ...940...30B} (the subsample containing 3,051 common sources). The completeness is defined as the multiplicity of this ratio and $C(\mathrm{SEGUE})$). The purity is defined as the proportion of real low $\log g$ stars in the 1,903 BHB candidates and shown in the middle panel of Figure~\ref{fig:compleAndPurity}. We can see that the completeness declines quickly as stars become fainter, while for the purity it shows little dependence on the magnitude and fluctuates randomly from 0.85 to 1.
  
\cite{2024ApJS..270...11J} propose a different method to estimate the purity of a BHB sample by counting the contribution of stars with $\mathrm{[M/H]}<-1$. Following their method, we cross-matched the XP spectrally confirmed BHB catalog with the LAMOST OBA-type stars catalog of \cite{2022A&A...662A..66X} and obtained 4,332 common sources. The LAMOST OBA-type stars catalog spans a broader range of magnitudes than the SEGUE BHB catalog, with $G$ spanning approximately from 4.5 to 21.5. Among these common sources, 3,837 stars are classified as metal-poor with $\mathrm{[Fe/H]_\mathrm{LAMOST}}<-1$ by \cite{2022A&A...662A..66X}, which also suggests a similar overall purity of $3,837/4,332\sim89\%$. We also divided these common sources into different bins to investigate the dependence of the purity on stellar color and magnitude. As shown in the bottom panel of Figure~\ref{fig:compleAndPurity}, the purity does not exhibit significant variation with $G_\mathrm{BP}-G_\mathrm{RP}$ or $G$, which is very similar to the purity defined by the proportion of low $\log g$ stars in the middle panel.

\subsection{Selection function}
The selection function of the BHB sample is defined as Eq~\ref{eq:SFall}:
\begin{equation}
\begin{aligned}
S_\mathrm{BHB}&=\frac{n_\mathrm{BHB;sub}(l,b,c,m)}{n_\mathrm{BHB;ph}(l,b,c,m)}\\&=
\frac{n_\mathrm{BHB;XP}(l,b,c,m)}{n_\mathrm{BHB;ph}(l,b,c,m)}\times\frac{n_\mathrm{BHB;sub}(l,b,c,m)}{n_\mathrm{BHB;XP}(l,b,c,m)},\label{eq:SFall}
\end{aligned}
\end{equation}
where $n_\mathrm{BHB;sub}(l,b,c,m)$, $n_\mathrm{BHB;XP}(l,b,c,m)$, and $n_\mathrm{BHB;ph}(l,b,c,m)$ are BHB star counts in our sample, Gaia XP spectra, and the complete photometric data in the color-magnitude diagram, respectively. We used the subscript `sub' to indicate that our BHB sample is only a subsample of BHB stars in the XP spectra. The first factor of $S_\mathrm{BHB}$ can be written as follows:
\begin{equation}
\begin{aligned}
\frac{n_\mathrm{BHB;XP}(l,b,c,m)}{n_\mathrm{BHB;ph}(l,b,c,m)}&=\frac{n_\mathrm{XP}(l,b,c,m)\times r_\mathrm{BHB;XP}(l,b,c,m)}{n_\mathrm{ph}(l,b,c,m)\times{r_\mathrm{BHB;ph}(l,b,c,m)}}\\&=S_\mathrm{XP}(l,b,c,m)\times\frac{r_\mathrm{BHB;XP}(l,b,c,m)}{r_\mathrm{BHB;ph}(l,b,c,m)}
,\label{eq:SFXP}
\end{aligned}
\end{equation}
where $n_\mathrm{XP}(l,b,c,m)$ and $n_\mathrm{ph}(l,b,c,m)$ are star counts in Gaia XP spectra and the complete photometric survey in the color-magnitude diagram. For stars in a certain $(l,b,c,m)$ bin, they might be BHB stars or some other kinds of stars like nearby Blue Stragglers. We defined $r_\mathrm{BHB;XP}(l,b,c,m)$ and $r_\mathrm{BHB;ph}(l,b,c,m)$ as the ratio of BHB stars to all stellar objects in Gaia XP spectra and a complete photometric survey. We assumed that the sample of Gaia XP spectra survey does not favor one type of star over another in a certain $(l,b,c,m)$ bin, which means $r_\mathrm{BHB;XP}(l,b,c,m)\approx r_\mathrm{BHB;ph}(l,b,c,m)$. We adopted the Gaia DR3 photometric survey as the reference since it is almost complete for stars of $G<19$ \citep{2023A&A...669A..55C}. Using the Python package gaiaunlimited \citep{2023A&A...677A..37C}, we generated the selection function $S_\mathrm{XP}(l,b,c,m)$ at the resolution of the HEALPix\footnote[1]{HEALPix, an acronym for Hierarchical Equal Area isoLatitude Pixelization of a sphere, is an algorithm that describes the Sky positions $(l,b)$. Stars with similar coordinates $(l,b)$ have the same HEALPix index number. Further information about HEALPix is available in the Gaia archive documentation.} level 5 in the Sky positions (12,288 equally sized areas, referred to as the 12,288 base pixels), $G\in[3, 20]$ in steps of 0.2, and $G-G_\mathrm{RP}\in[-0.05, 0.45]$ in bins of 0.25.

The second factor ${n_\mathrm{BHB;sub}(l,b,c,m)}/{n_\mathrm{BHB;XP}(l,b,c,m)}$ is the probability $P_\mathrm{BHB}$ that a BHB star can be correctly selected from XP spectra. $n_\mathrm{BHB;XP}$ is an unknown component because we cannot identify all BHB stars from XP spectra. However, a fraction of them have been identified by the SEGUE data, which are the common 3,051 sources mentioned above, and $P_\mathrm{BHB}$ estimated from this subsample is used as an approximation of the second factor. The small number of common sources prevents us from exploring $P_\mathrm{BHB}$ for different sky positions. Assuming that $P_\mathrm{BHB}$ is irrelevant to the positions in the sky, it corresponds only to the completeness mentioned above in Figure~\ref{fig:compleAndPurity}. We ignored the effect of stellar color since it is only a secondary determinant of the completeness. We assumed C(SEGUE) as a constant since it should not vary significantly with magnitude. For stars of $G\geq14$, the probability $P_\mathrm{BHB}$ was obtained by performing a quadratic 1-D interpolation to the dashed black line in the top panel of Figure~\ref{fig:compleAndPurity}. The completeness increases as BHB stars become brighter and approaches $1$ for stars of $G=14.2$. Therefore, for stars brighter than $G=14.2$ we assumed a probability $P_\mathrm{BHB}=1$. The selection function $S_\mathrm{BHB}$ is defined as: 
\begin{equation}
    S_\mathrm{BHB}(l,b,G-G_\mathrm{RP},G)=S_\mathrm{XP}(l,b,G-G_\mathrm{RP},G)P_\mathrm{BHB}(G),
\end{equation}
where $c$ is the stellar color of $G-G_\mathrm{RP}$, and $m$ is the apparent magnitude of $G$. The selection function is also provided as a parameter in our catalog.

\section{Density shape of the inner stellar halo}\label{subsec:density}
Understanding the spatial distribution of halo stars, specifically the radial density profile and the vertical flattening, is critical to disentangling the mass assembly history of the Milky Way. The radial density profile of the stellar halo is typically described by a broken power law (BPL) with a constant flattening $q\sim0.7$ in previous studies \citep[e.g.,][]{2009MNRAS.398.1757W,2011MNRAS.416.2903D,2014ApJ...788..105F,2015ApJ...809..144X,2015A&A...579A..38P,2016MNRAS.463.3169D,2024MNRAS.531.4762M}. A break radius $r_\mathrm{break}$ has been identified between 15 and 30 kpc. The radial density profile inside $r_\mathrm{break}$ follows a power law with an index of about $2.5-3$, while stars outside $r_\mathrm{break}$ follow a much steeper one with an index of approximately $4-5$. Some studies suggest that there might be two break radii at the inner ($r_\mathrm{break}\sim10-15$ kpc) and outer ($r_\mathrm{break}\sim20-30$) parts of the stellar halo \citep{2022AJ....164..241Y,2022AJ....164..249H}. Recently, several studies find that $r_\mathrm{break}$ is not necessarily needed assuming a varying flattening, in which the radial density profile can be described by a single power law (SPL) with an index of $4-5$ using an oblate ellipsoid model \citep{2016MNRAS.463.3169D,2018ApJ...859...31H,2018MNRAS.473.1244X,2022AJ....164...41W,2023MNRAS.525.3075C} or an index of 2.96 using a triaxial ellipsoid model \citep{2018MNRAS.474.2142I}. 

\cite{2017RAA....17...96L} and \cite{2018MNRAS.473.1244X} describe a method to derive the stellar density profile along each line of sight ($\textit{l}-\textit{b}$ plate) for the LAMOST spectroscopic survey. Using their method, we can explore the radial density profile of the stellar halo with varying flattening through our identified BHB sample. In their works, the ratio between the stellar density obtained from the photometric ($\nu_\mathrm{ph}$) and spectroscopic ($\nu_\mathrm{sp}$) surveys is defined as:
\begin{equation}
    \nu_\mathrm{ph}(D|\textit{l}, \textit{b}, \textit{c}, \textit{m}) = \nu_\mathrm{sp}(D|\textit{l}, \textit{b}, \textit{c}, \textit{m}){S^{-1}(\textit{l}, \textit{b}, \textit{c}, \textit{m})},
    \label{eq:nu}
\end{equation}
where $D$ is the distance along a given line of sight, and $S$ is the selection function of BHB stars ($S_\mathrm{BHB}$) in this study.  

After integrating over $\textit{c}$ and $\textit{m}$, the stellar density profile for a given line of sight is defined as:
\begin{equation}
\nu_\mathrm{ph}(D|l, b) = \iint{\nu_\mathrm{sp}(D|\textit{l}, \textit{b}, \textit{c}, \textit{m})S^{-1}(\textit{l}, \textit{b}, \textit{c}, \textit{m})\mathrm{d}c\mathrm{d}m}.
\label{eq:ph}
\end{equation}

To account for errors arising from distance, we employed a kernel density estimation (KDE) method to derive $\nu_\mathrm{sp}$ along a given line of sight. The contribution of a star $i$ is treated as a probability density function $p_i$ extending along the corresponding line of sight, and $p_i$ is defined as:
\begin{equation}
	p_i(D) = \frac{\mathcal{N}(D|D_i, \sigma_{D_i}^2)}{\int_{D_\mathrm{min}}^{D_\mathrm{max}}\mathcal{N}(D_x|D_i, \sigma_{D_i}^2)dD_x},
	\label{eq:pi}
\end{equation}
where $\mathcal{N}$ is a normal function. $D_i$ is the estimated distance of star $i$, and $\sigma_{D_i}$ is the uncertainty of the estimated distance. In this study, $\sigma_{D_i}$ is defined as:
\begin{equation}
    \sigma_{D_i} = \frac{D_i(84\%)-D_i(16\%)}{2}
\end{equation}
where $D_i(16\%)$ and $D_i(84\%)$ are the $16\%$ and $84\%$ distributions of $d_\mathrm{helio}$ obtained by substituting $M_{G}(P_{16})$ and $M_{G}(P_{84})$ into Equation~\ref{eq:fitdistance}. We setted $D_\mathrm{min} = 0$ kpc and $D_\mathrm{max} = 200$ kpc.

We took into account the contribution of all stars around $\textit{c}$ and $\textit{m}$. The density profile $\nu_\mathrm{sp}$ is defined as:
\begin{equation}
\nu_\mathrm{sp}(D|\textit{l}, \textit{b}, \textit{c}, \textit{m}) = \frac{1}{{\Omega}D^2}\sum_{i}^{n_\mathrm{sp}(\textit{l}, \textit{b}, \textit{c}, \textit{m})}p_i(D),
\label{eq:sp}
\end{equation}
where $\Omega$ is the solid angle of the given line of sight, and $n_\mathrm{sp}(\textit{l}, \textit{b},\textit{c}, \textit{m})$ is the number of spectroscopic stars within a given $(\textit{l}, \textit{b},\textit{c}, \textit{m})$ bin. Note that in this form, the derived $\nu_\mathrm{ph}$ is a continuous function of $D$.

In the LAMOST survey, stars are observed in thousands of plates (lines of sight) with a constant solid angle. To achieve a similar situation, we divided the sky positions into 3,072 plates at the resolution of the HEALPix level 4, and BHB stars (potential globular cluster members are removed) were distributed in different lines of sight according to their Galactic coordinates. 
Since $\Omega, \mathrm{d}c, \mathrm{d}m$, and C(SEGUE) are constant values that have no impact on the obtained density shape, we normalized them to 1 for computational convenience. $\nu_\mathrm{ph}$ is obtained by combing Equation~\ref{eq:ph} and ~\ref{eq:sp}. We only adopted $\nu_\mathrm{ph}$ at the positions where we can find a star, since the stellar density derived at the positions without sampled stars may suffer from large uncertainties. In other words, every BHB star in our sample is assigned a $\nu_\mathrm{ph}$ that represents an estimate of the stellar density at its position.

\begin{figure*}
    \centering
    \includegraphics[width = \textwidth]{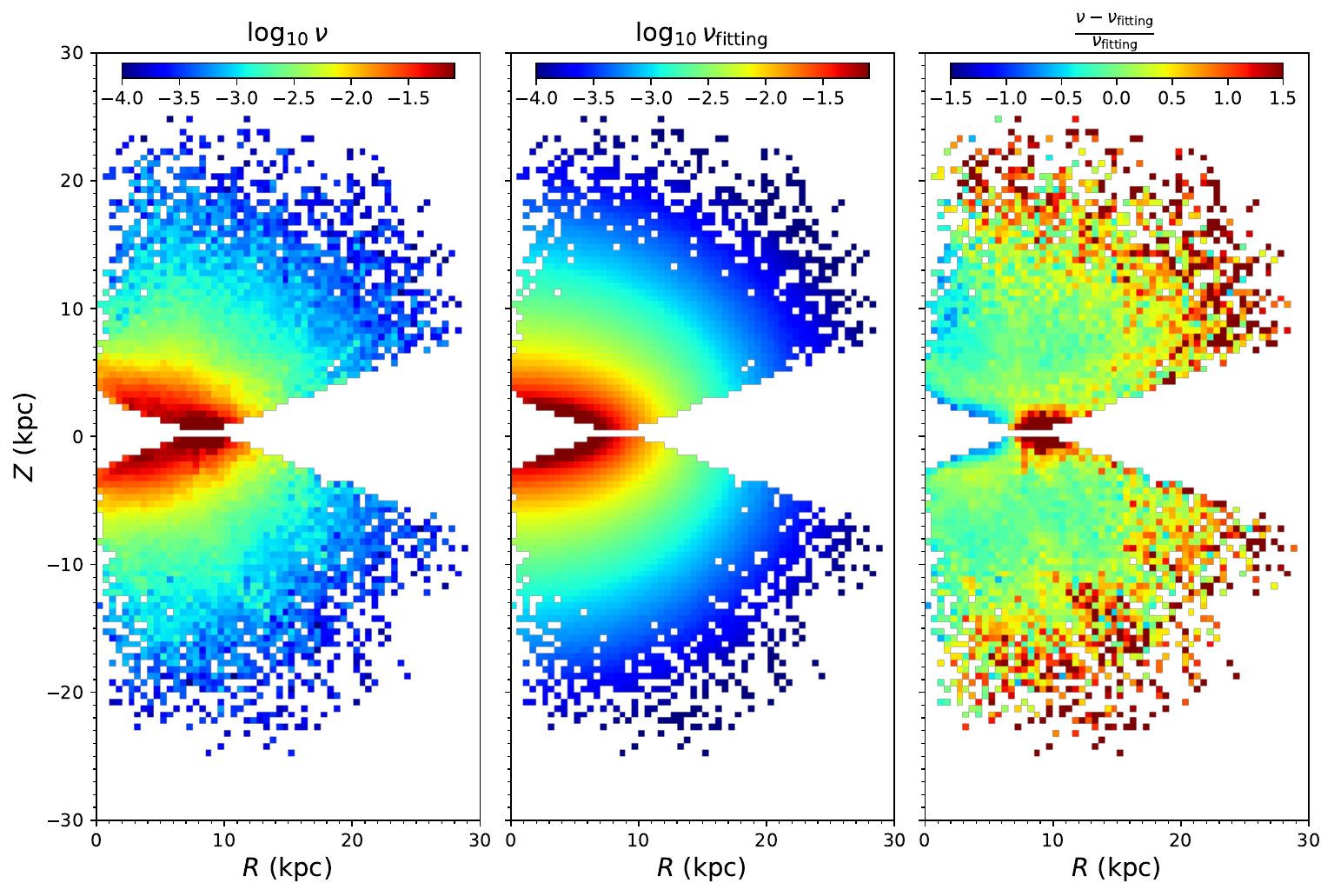}
    \caption{Number density map of the stellar halo in $R-Z$ space for the BHB data ($\log_{10}\nu$, left panel) and the model ($\log_{10}\nu_\mathrm{fitting}$, middle panel). The pixel size is $0.5\times0.5$ (kpc). The color in each pixel represents the mean density on a logarithmic scale. The right panel shows the residual map ($(\nu-\nu_\mathrm{fitting})/\nu_\mathrm{fitting}$), where the most prominent difference is seen in the solar neighborhood and the outer region ($r_\mathrm{gc}>15\,\mathrm{kpc}$).}
    \label{fig:median}
\end{figure*}

Figure~\ref{fig:median} shows the median value of $v_\mathrm{ph}$ in the $|R|-Z$ pixel map. $R$ and $Z$ are cylindrical Galactocentric coordinates obtained from the conventions in astropy \citep{2018AJ....156..123A}. We used the default values of the Solar Galactocentric distance $r_{\mathrm{gc},\odot}$ = 8.122 kpc \citep{2018A&A...615L..15G}, and height $Z_{\odot}$ = 20.8 pc \citep{2019MNRAS.482.1417B}. We can clearly see a vertically flattened shape of the stellar halo that shows a larger flattening at a smaller radius. To characterize the vertical flattening quantitatively, we fit the density profile as a function of the flattened Galactic radius $r$ with an ellipsoid model as follows:
\begin{align}
	&\nu(r) = {\nu_0}r^{-\alpha},\label{eq:spl}\\
	&r = \sqrt{R^2+(Z/q(r))^2},
\label{eq:eps}
\end{align}
where $q(r)$ is the flattening parameter.

As noted by \citet{2018MNRAS.473.1244X}, constraining $r$ and $q$ is challenging in the Galactocentric cylindrical coordinate because $q$ is very sensitive to the slope of the iso-density surface, which becomes too steep at low latitudes. Therefore, following \citet{2018MNRAS.473.1244X} and \citet{2022ApJ...924...23W}, we will conduct our analysis in the Galactocentric polar coordinates where $R$ and $Z$ can be defined as follows:
\begin{align}
	& R = r\,\mathrm{cos}(\eta), \label{eq:R} \\
	& Z = r\,q\,\mathrm{sin}(\eta), \label{eq:z}\\
	&r_\mathrm{gc} = \sqrt{R^2+Z^2} \label{eq:rgc}\\
	&\mathrm{sin}(\theta) = Z/r_\mathrm{gc}. \label{eq:theta}
\end{align}

Combining Equation~\ref{eq:R}, ~\ref{eq:z}, and ~\ref{eq:rgc}, we rewrite $r_\mathrm{gc}$ as:
\begin{equation}
	r_\mathrm{gc} = r\sqrt{1 + (q^2 - 1)\mathrm{sin}^2(\eta)}.
	\label{eq:rgc-q}
\end{equation}

Using Equation~\ref{eq:theta}, we write $\sin(\eta)$ as: 
\begin{equation}
	\mathrm{sin}(\eta) = \frac{r_\mathrm{gc}\,\mathrm{sin}(\theta)}{rq}.
	\label{eq:eta}
\end{equation} 
 
Substituting Equation~\ref{eq:eta} into Equation~\ref{eq:rgc-q}, we can fit the iso-density contour in $r_\mathrm{gc}-\sin(\theta)$ space with Equation~\ref{eq:final},
\begin{equation}
	r_\mathrm{gc} = r q\sqrt{\frac{1}{q^2-(q^2-1)\sin^2(\theta)}}.
	\label{eq:final}
\end{equation}

\begin{figure*}
    \centering
    \includegraphics[width = 0.8\textwidth]{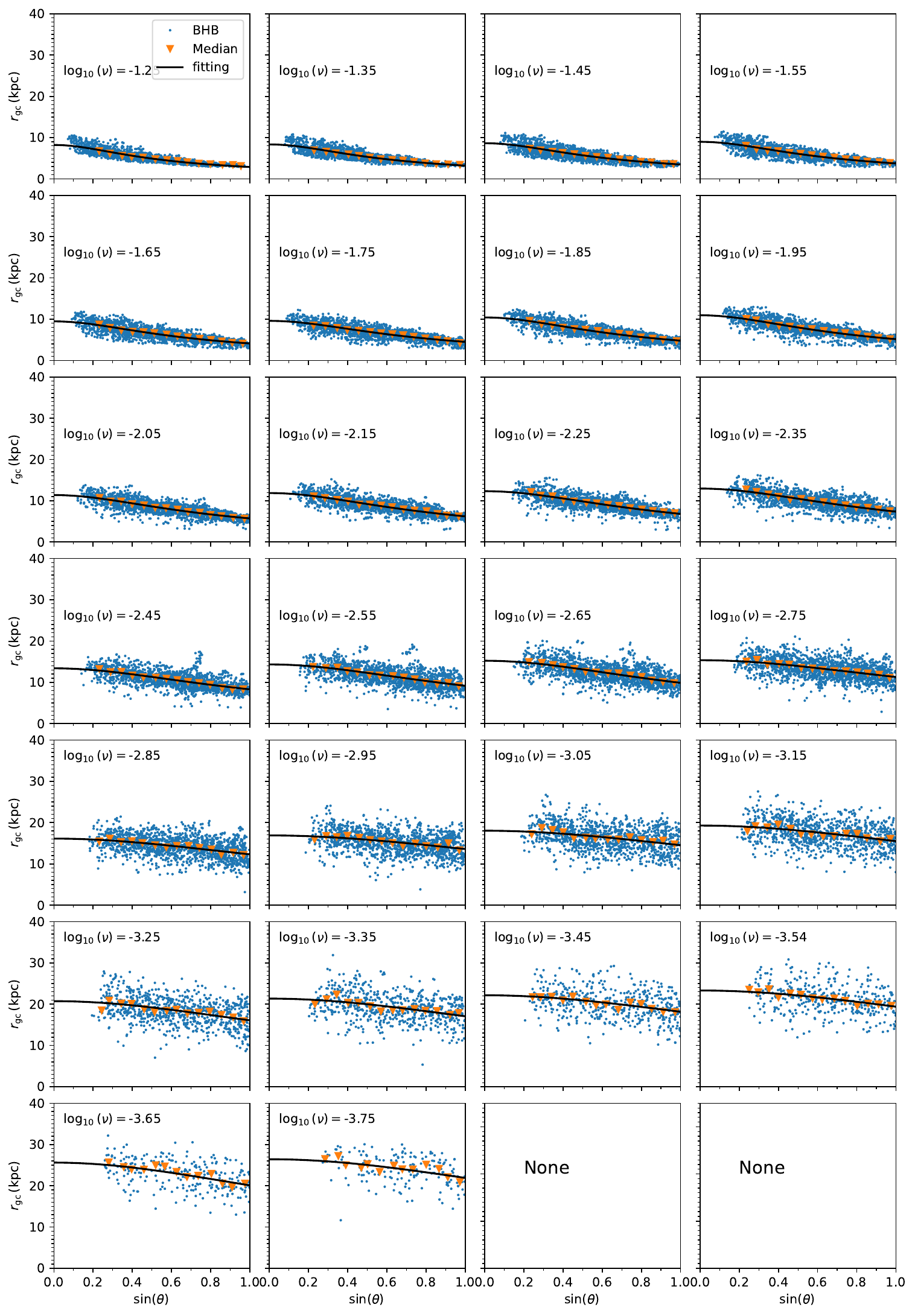}
    \caption{Fitting results of all BHB stars in different bins of $\log_{10}(\nu)$. The blue points are BHB stars in a certain bin and the orange inverted triangles represent the median value of $r_\mathrm{gc}$. The black lines are the best fitting results with the specific parameters recorded in Table~\ref{tab:fitting}.}
    \label{fig:fitting}
\end{figure*}

\begin{figure}
    \centering
    \includegraphics[width = 8.8cm]{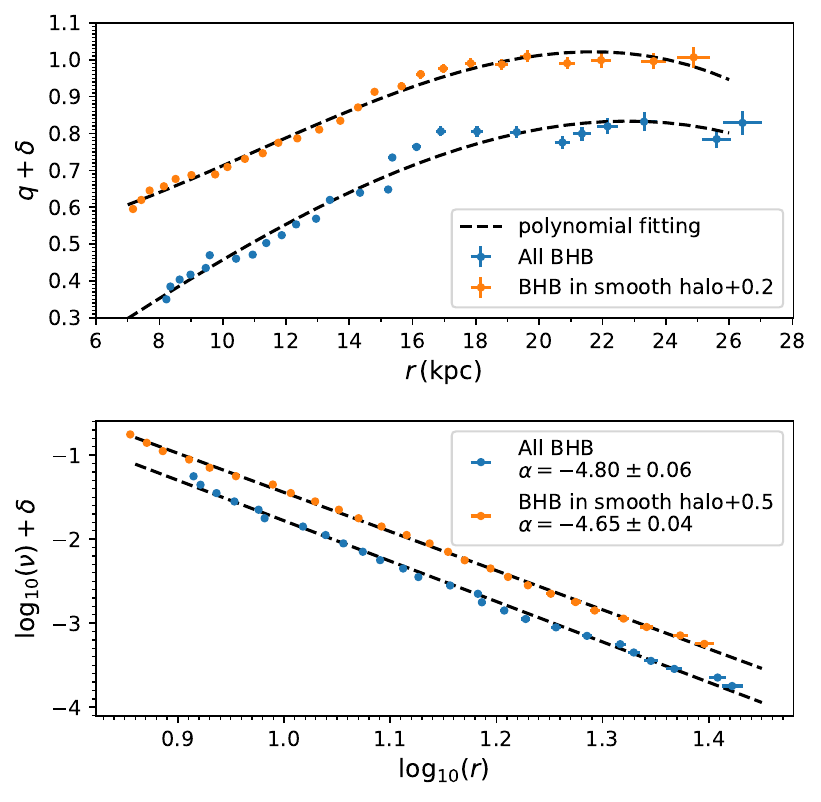}
    \caption{Density shape and profile of the stellar halo traced by the BHB sample. The top panel shows the variation in $q$ with increasing $r$ for all BHB stars (blue points) and the subsample (orange points). To better distinguish the results of the two samples, a constant of 0.2 is added to $q$ of the subsample. The stellar halo is largely vertically flattened in the inner part and transformed into a nearly spherical shape at the outer radii. The bottom panel shows the relationship between the radial stellar density $\log(\nu)$ and the flattened radius $r$. Assuming a variable flattening, the radial density profile of the stellar halo derived from all BHB stars (blue points) can be well described by a SPL with an index $\alpha=-4.80\pm0.06$, and for the subsample (orange points, a constant of 0.5 is added to $\log(\nu)$) the best result is $\alpha=-4.65\pm0.04$.}
    \label{fig:densityANDradius}
\end{figure}

We divided the BHB stars into different bins of $\log_{10}(\nu)$ with a bin size of 0.1 in the range $-1.2<\log_{10}(\nu_\mathrm{ph})<-3.8$. We fit the spatial distribution ($r_\mathrm{gc},\sin(\theta)$) of the BHB stars with an ellipse of Equation~\ref{eq:final} in each $\log_{10}(\nu)$ bin by scipy.optimize.curve\_fit. Our fitting results for all bins are shown in Figure~\ref{fig:fitting} and the specific values are recorded in Table~\ref{tab:fitting}. The variation in $q$ with increasing $r$ shows a clear tendency that the stellar halo becomes more vertically flattened in the inner part. In Figure~\ref{fig:densityANDradius}, we find that the radial density profile of the stellar halo can be simply described by a SPL of an index of $\alpha=-4.80\pm0.05$ with varying flattening. Our result is almost the same as the radial density profile ($\alpha=-5.03\pm0.64$) derived from the LAMOST K-giant sample in \cite{2018MNRAS.473.1244X}, but has a smaller dispersion that might be related to the larger dataset and the accurate distance estimation of the BHB sample.

\begin{table}
\caption{Best estimated parameters of the stellar halo shape derived from the BHB catalog.}              
\label{tab:fitting}      
\centering                                    
\begin{tabular}{ccccccc}          
\hline\hline                        
$\log_{10} (\nu)$ & $\mathrm{N}^a$ & ${r}^a$ & $q^a$ & $\mathrm{N}^b$ & ${r}^b$ & $q^b$\\    
\hline                                   
-1.25 & 1,701 & 8.22  & 0.35 &  1,220 & 7.17  & 0.40 \\
-1.35 & 2,023 & 8.35  & 0.38 &  1,527 & 7.43  & 0.42 \\
-1.45 & 2,561 & 8.64  & 0.40 &  2,078 & 7.69  & 0.45 \\
-1.55 & 2,500 & 8.98  & 0.42 &  2,217 & 8.14  & 0.46 \\
-1.65 & 2,469 & 9.46  & 0.44 &  2,189 & 8.51  & 0.48 \\
-1.75 & 2,350 & 9.59  & 0.47 &  2,224 & 9.01  & 0.49 \\
-1.85 & 2,384 & 10.42 & 0.46 & 2,257 & 9.76  & 0.49 \\
-1.95 & 2,212 & 10.95 & 0.47 & 2,097 & 10.15 & 0.51 \\
-2.05 & 2,177 & 11.38 & 0.50 & 2,070 & 10.70 & 0.53 \\
-2.15 & 2,046 & 11.86 & 0.52 & 1,919 & 11.26 & 0.55 \\
-2.25 & 1,987 & 12.32 & 0.55 & 1,881 & 11.76 & 0.57 \\
-2.35 & 2,028 & 12.95 & 0.57 & 1,887 & 12.36 & 0.59 \\
-2.45 & 1,853 & 13.39 & 0.62 & 1,707 & 13.05 & 0.61 \\
-2.55 & 1,808 & 14.34 & 0.64 & 1,593 & 13.72 & 0.63 \\
-2.65 & 1,818 & 15.23 & 0.65 & 1,608 & 14.27 & 0.67 \\
-2.75 & 1,727 & 15.36 & 0.73 & 1,453 & 14.80 & 0.71 \\
-2.85 & 1,477 & 16.12 & 0.76 & 1,249 & 15.65 & 0.73 \\
-2.95 & 1,356 & 16.89 & 0.81 & 1,091 & 16.25 & 0.76 \\
-3.05 & 1,194 & 18.04 & 0.81 & 937  & 16.97 & 0.78 \\
-3.15 & 1,018 & 19.29 & 0.80 & 789  & 17.83 & 0.79 \\
-3.25 & 832  & 20.73 & 0.78 & 650  & 18.82 & 0.79 \\
-3.35 & 654  & 21.35 & 0.80 & 549  & 19.62 & 0.81 \\
-3.45 & 496  & 22.16 & 0.82 & 439  & 20.88 & 0.79 \\
-3.54 & 387  & 23.31 & 0.83 & 328  & 21.95 & 0.80 \\
-3.65 & 263  & 25.61 & 0.78 & 224  & 23.61 & 0.80 \\
-3.75 & 182  & 26.42 & 0.83 & 147  & 24.88 & 0.81 \\
\hline
\end{tabular} 
\footnotesize
\begin{itemize}
\item N is the number of stars in the bin, $^a$ indicates the BHB sample, and $^b$ indicates the BHB subsample of the smooth stellar halo.
\end{itemize}
\end{table}

\begin{figure}
    \centering
    \includegraphics[width = 8.8cm]{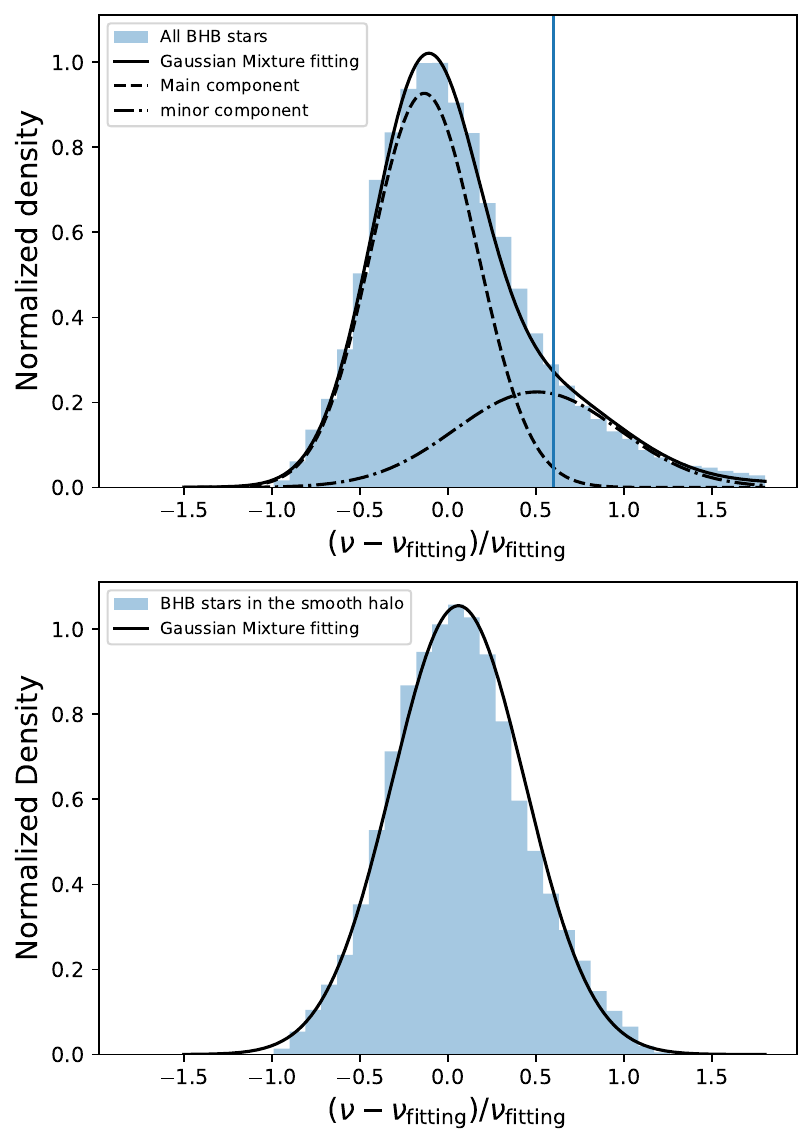}
    \caption{Normalized density distribution of $(\nu-\nu_\mathrm{fitting})/\nu_\mathrm{fitting}$ obtained from the whole BHB stars (top panel) and the subsample of the smooth halo (bottom panel). The blue vertical line represents the position of $(\nu-\nu_\mathrm{fitting})/\nu_\mathrm{fitting}=0.6$.}
    \label{fig:histdist}
\end{figure}

The number density map ($\nu$) and the model ($\nu_\mathrm{fitting}$) fit are shown in Figure~\ref{fig:median}. In the spatial positions of each BHB star, $\nu_\mathrm{fitting}$ was generated based on parameters derived from the binned data in Figure~\ref{fig:fitting} and ~\ref{fig:densityANDradius}. We applied a second-order polynomial function ($q=f(r)$) to describe the $r-q$ relationship and $q$ is extrapolated as a constant of 0.81 beyond $r_{q}=23\,\mathrm{kpc}$. We can see that the number density map of the BHB sample is generally consistent with the model except for some prominent overdensities in the solar neighborhood and the outer region ($r_\mathrm{gc}>15\,\mathrm{kpc}$). From Figure~\ref{fig:histdist}, we find that the density distribution of $\frac{\nu-\nu_\mathrm{fitting}}{\nu_\mathrm{fitting}}$ can be well described by a Gaussian mixture model of two components, of which the primary is centered at -0.12 with a standard deviation of 0.30, and the minor is centered at around 0.53 with a standard deviation of 0.50. BHB stars in the minor component shows an underestimated $\nu_\mathrm{fitting}$, which is likely to be related to the prominent overdensities found in the residual map.

Several contributing factors may account for the presence of the minor component. The selection function correction may not be suitable for all BHB stars due to the assumptions made in section~\ref{subsec:sf}. The error in the distance estimation ($\sigma_{D_i}$) increases for more distant BHB stars, potentially leading to an increased scatter relative to the fitting line for stars in the outer halo bins of Figure~\ref{fig:median}. Considering the existence of multiple stellar substructures in the Galactic halo, it is natural that a fraction of our selected BHB stars belong to some unresolved substructures. The presence of non-axisymmetric substructures may introduce a bias to the fitting, since the stellar halo is viewed as a smooth component in our model. The inclusion of contaminants can cause an overestimation of $\nu$, and may explain the most prominent overdensity in the solar neighborhood considering the large numbers of metal-rich stars in this region. Recent results from nearby stars suggest an early formation of the disk component within the first few billion years \citep{2022MNRAS.514..689B,2022arXiv220402989C,2024MNRAS.533..889Z}, and a disk-like geometry could exist even for a very old stellar population of $\sim13$ Gyr \citep{2025NatAs...9..101X}. It is hard for us to remove all disk stars due to the lack of metallicity and kinematic information. Therefore, it is possible that a small fraction of our identified BHB stars with low $|Z|$ actually belong to the old disk.

\begin{figure}
    \centering
    \includegraphics[width = 8.8cm]{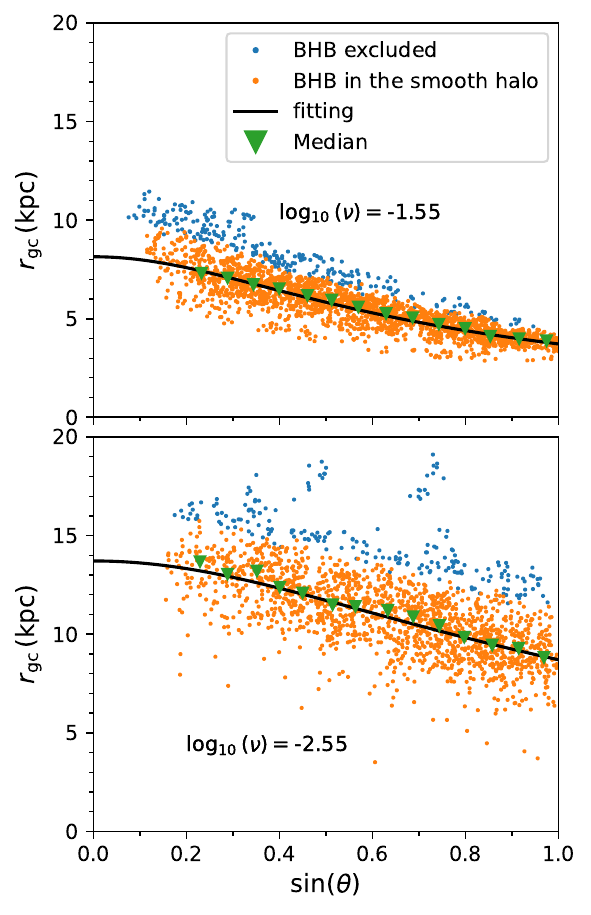}
    \caption{Fitting results in two bins of $\log_{10}(\nu)=-1.55$ (top panel) and $-2.55$ (bottom panel) for the BHB subsample of the smooth halo. In each panel, the orange points are BHB stars within the specified bin, while the green inverted triangles indicate the median value of $r_\mathrm{gc}$. The black lines depict the best fitting results of the subsample. The blue points denote stars that were previously included in Figure~\ref{fig:fitting} but have been excluded here due to their large residual of $\frac{\nu-\nu_\mathrm{fitting}}{\nu_\mathrm{fitting}}\geq0.6$. In the top panel, the excluded BHB stars are responsible for the prominent overdensity in the solar neighborhood in Figure~\ref{fig:median}. In the bottom panel, these excluded stars are likely associated with the overdensity in the outer regions of the halo.}
    \label{fig:fitting_constant}
\end{figure}

\begin{figure*}
    \centering
    \includegraphics[width = \textwidth]{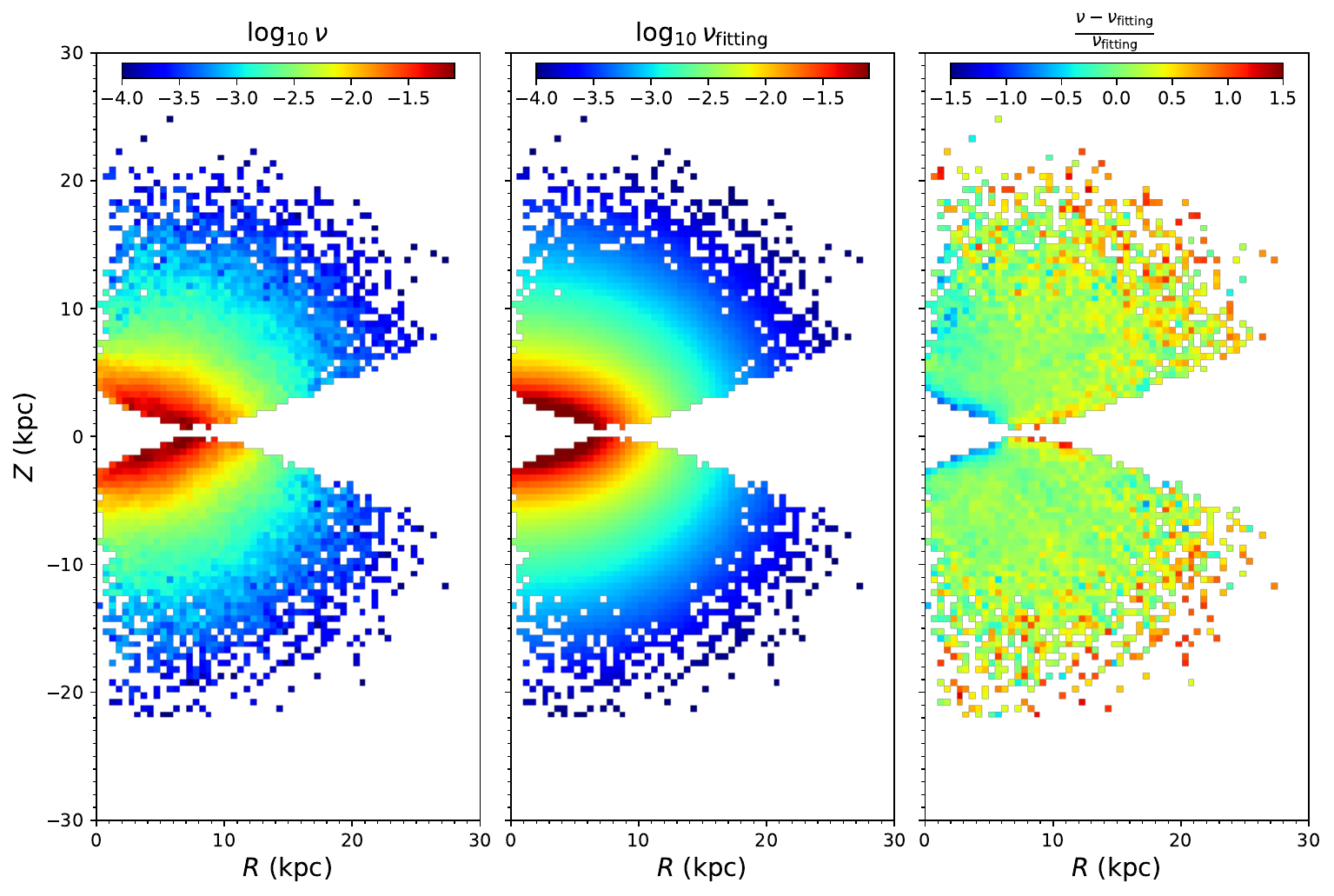}
    \caption{Number density map of the stellar halo in $R-Z$ space for the BHB stars in the subsample ($\log_{10}\nu$, left panel) and the model ($\log_{10}\nu_\mathrm{fitting}$, middle panel).The pixel size is $0.5\times0.5$ (kpc). The color in each pixel represents the mean density on a logarithmic scale. The residual map ($(\nu-\nu_\mathrm{fitting})/\nu_\mathrm{fitting}$) presented in the right panel indicates a good consistency between the data and the model, where no prominent overdensity exists in the solar neighborhood and the outer parts of the stellar halo.}
    \label{fig:median_constant}
\end{figure*}

To reduce the influence of the minor component, we applied a rough cut of $\frac{\nu-\nu_\mathrm{fitting}}{\nu_\mathrm{fitting}}<0.6$ and defined the remaining stars as the subsample of the smooth stellar halo. We then repeated the fitting process and presented the results for two representative bins in Figure~\ref{fig:fitting_constant}. As is shown in Figure~\ref{fig:densityANDradius}, the main tendency that the stellar halo is more vertically flattened in the inner radii remains unchanged for the subsample, but the $r-q$ relationship is now better described by a second-order polynomial function without evident outliers at $r\sim17\,\mathrm{kpc}$. Overall, the increasing trend of ${q}$ with $r$ is smoother and more consistent for the subsample. The density profile exhibits only a slight change and is well characterized by a SPL of an index of $\alpha=-4.65\pm0.04$. For each BHB star in the subsample, we generate its $\nu_\mathrm{fitting}$ using the updated $r-q$ relationship and density profile, where $q$ is extrapolated as a constant of 0.80 beyond $r_{q}=23\,\mathrm{kpc}$. In Figure~\ref{fig:histdist}, the density distribution of $\frac{\nu-\nu_\mathrm{fitting}}{\nu_\mathrm{fitting}}$ for the smooth halo is well fit by a single Gaussian function, with a mean of 0.05 and a standard deviation of 0.38. A significant improvement in consistency between the BHB data and the model is clearly shown in the residual map of Figure~\ref{fig:median_constant}. The most prominent overdensity in the solar neighborhood disappears, and the discrepancies in the outer part of the stellar halo are also largely improved.

\section{Summary}\label{sec:summary}
In this study, we constructed a catalog of 44,524 BHB stars selected from the Gaia DR3 XP spectra by combining photometric and spectroscopic data. Using a fraction of BHB stars with a reliable parallax, we derived a relationship between the absolute magnitude $M_{G}$ and the dereddened stellar color $G_\mathrm{BP;0}-G_\mathrm{RP;0}$. A purity of around $90\%$ was obtained by adopting the SEGUE spectroscopic-confirmed BHB catalog of \cite{2022ApJ...940...30B} and the hot stars catalog of \cite{2022A&A...662A..66X} as the reference datasets. We estimated the selection function of these BHB stars by comparing our catalog to the complete photometric survey of Gaia DR3. Assuming that the density shape varies with the radius, the radial density profile of the Galactic stellar halo traced by the BHB catalog can be well defined by a single power law with $\alpha=-4.80\pm0.06$. A comparison between the BHB data ($\nu$) and model ($\nu_\mathrm{fitting}$) suggests a possible bias introduced by contaminants or stars from the stellar substructures. After removing these obvious outliers ($\frac{\nu-\nu_\mathrm{fitting}}{\nu_\mathrm{fitting}}>0.6$), we obtained a smoother and more consistently positive relationship of $r-q$, and the density profile only shows a slight change with the best-estimated index $\alpha=-4.65\pm0.04$.  

This study is an attempt to construct a full-sky covered complete BHB catalog of the Milky Way from the low-resolution XP spectra. Although the low-resolution XP spectra prevent us from obtaining an extremely accurate estimation of the stellar parameters, we are still able to achieve a purity of around $87\%$, similar to the BHB catalog of \cite{2024ApJS..270...11J}, which was selected from the higher resolution spectra of LAMOST. However, the completeness, which drops from about 90\% at $G=14$ mag to around 40\% at $G=17$ mag, suggests that there are still a large number of BHB stars that wait to be identified from the XP spectra, especially in the faint magnitudes. The lack of radial velocity measurements in this BHB catalog makes exploring the phase-space distribution of the stellar halo difficult. The upcoming Chinese Space Station Telescope (CSST) spectroscopic survey will release hundreds of millions of high-quality low-resolution ($R>200$) slitless spectra. We believe that our selection method will also be appropriate for the CSST low-resolution spectra survey (a limiting magnitude of around 23 mag in U, V, and I bands), which is much deeper than the Gaia XP spectra (a limiting magnitude of around 17.5 mag in G band) and will allow us to explore the outer regions of the stellar halo.

\section{Data availability}
Tables of the selected BHB members are only available in electronic form at the CDS via anonymous ftp to cdsarc.u-strasbg.fr (130.79.128.5) or via http://cdsweb.u-strasbg.fr/cgi-bin/qcat?J/A+A/.  

\begin{acknowledgements}
This study is supported by the National Key R\&D Program of China under grant Nos. 2023YFE0107800, 2024YFA1611900, and the National Natural Science Foundation of China under grant Nos. 12588202, 12273055, 12373020. This study is also supported by International Partnership Program of Chinese Academy of Sciences. Grant No. 178GJHZ2022040GC, and CAS Project for Young Scientists in Basic Research grant No. YSBR-062 and YSBR-092. We also thank the support by the China Manned Space Program with grant nos. CMS-CSST-2025-A11 and CMS-CSST-2025-A12. Xianhao Ye and Wenbo Wu acknowledge the support from the China Scholarship Council. CAP, DA, JIGH, and RRL acknowledge financial support from the Spanish Ministry of Science, Innovation and Universities (MICIU) projects PID2020-117493GB-I00 and PID2023-149982NB-I00. This research made use of computing time available on the high-performance computing systems at the Instituto de Astrofisica de Canarias. This work presents results from the European Space Agency (ESA) space mission Gaia. Gaia data are being processed by the Gaia Data Processing and Analysis Consortium (DPAC). Funding for the DPAC is provided by national institutions, in particular the institutions participating in the Gaia MultiLateral Agreement (MLA). The Gaia mission website is https://www.cosmos.esa.int/gaia. The Gaia archive website is https://archives.esac.esa.int/gaia. This job has made use of the Python package GaiaXPy, developed and maintained by members of the Gaia Data Processing and Analysis Consortium (DPAC), and in particular, Coordination Unit 5 (CU5), and the Data Processing Centre located at the Institute of Astronomy, Cambridge, UK (DPCI). Funding for SDSS-III has been provided by the Alfred P. Sloan Foundation, the Participating Institutions, the National Science Foundation, and the U.S. Department of Energy Office of Science. The SDSS-III web site is http://www.sdss3.org/. SDSS-III is managed by the Astrophysical Research Consortium for the Participating Institutions of the SDSS-III Collaboration including the University of Arizona, the Brazilian Participation Group, Brookhaven National Laboratory, Carnegie Mellon University, University of Florida, the French Participation Group, the German Participation Group, Harvard University, the Instituto de Astrofisica de Canarias, the Michigan State/Notre Dame/JINA Participation Group, Johns Hopkins University, Lawrence Berkeley National Laboratory, Max Planck Institute for Astrophysics, Max Planck Institute for Extraterrestrial Physics, New Mexico State University, New York University, Ohio State University, Pennsylvania State University, University of Portsmouth, Princeton University, the Spanish Participation Group, University of Tokyo, University of Utah, Vanderbilt University, University of Virginia, University of Washington, and Yale University. 

\end{acknowledgements}
\bibliographystyle{aa}
\bibliography{bibliography}
\begin{appendix}
\onecolumn
\section{Necessity of a combined photometric and spectroscopic cut}
\begin{figure*}[h!]
    \centering
    \includegraphics[width = 12cm,clip]{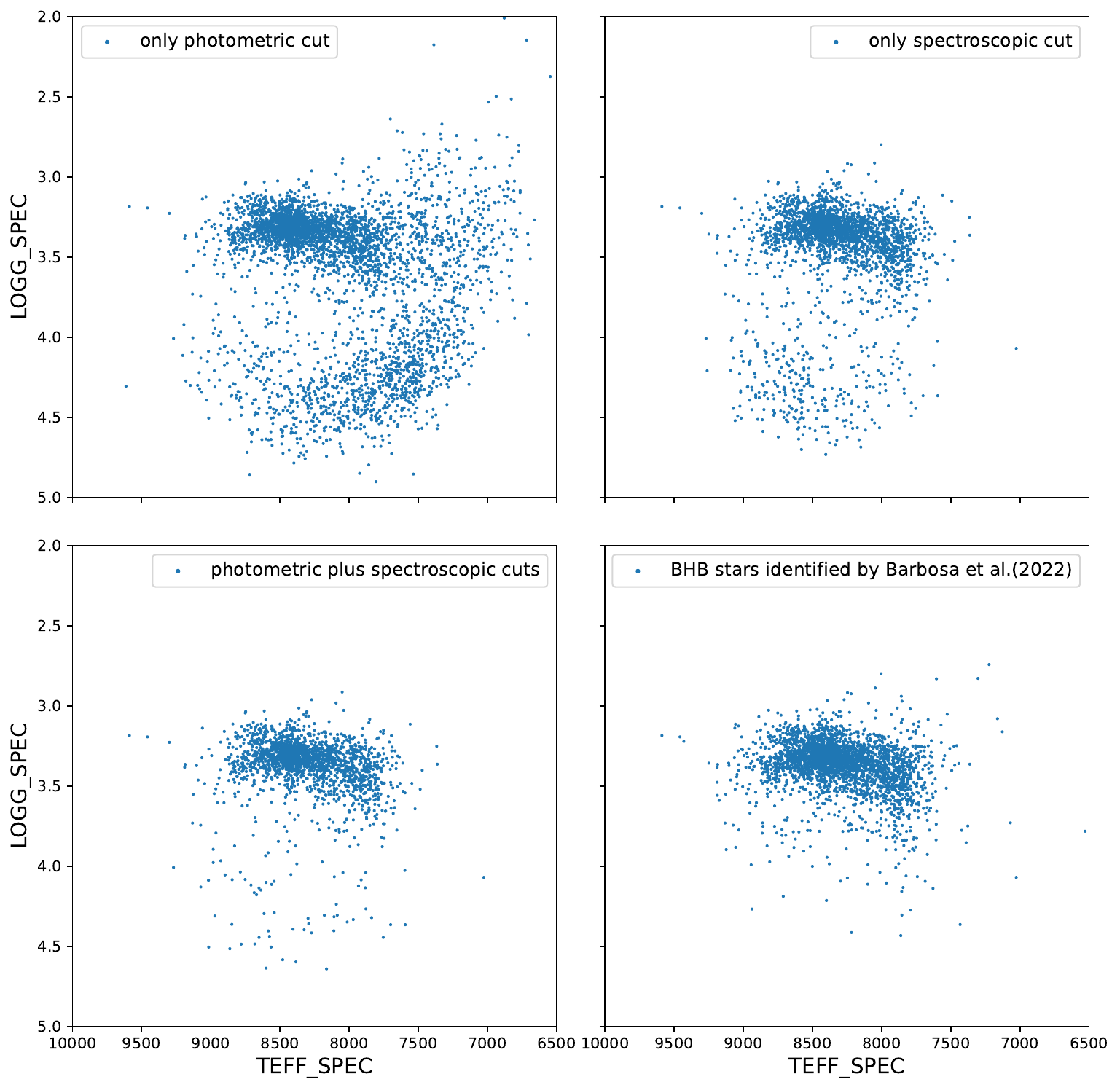}
    \caption{TEFF\_SPEC$-$LOGG\_SPEC diagram of the selected BHB candidates obtained by different cuts, and the BHB sample of \cite{2022ApJ...940...30B} is used as a reference. The proportion of contaminants from high surface gravity (LOGG\_SPEC$>$3.7) stars is 28\% for only photometric cuts, 14\% for only spectroscopic cuts, and 5\% for combined photometric/spectroscopic cuts.}
    \label{fig:differentcuts}
\end{figure*}

In this appendix, we conducted a test by applying only photometric cuts, only spectroscopic cuts, and combined photometric/spectroscopic cuts to Gaia XP spectra. Several high-latitude ($|b|>20^\circ$) selected BHB candidates are also available in the SEGUE spectroscopic catalog. Figure~\ref{fig:differentcuts} shows their distributions in the TEFF\_SPEC$-$LOGG\_SPEC diagram ($T_\mathrm{eff}-\log g$ provided by the SEGUE survey). Spectroscopic cuts perform better in the selection than photometric cuts, but there are still some contaminants that can only be removed by the photometric cuts. The combined photometric/spectroscopic cuts produce the cleanest sample, as indicated by the smallest rate (around 5\%) of contaminants from high surface gravity stars (LOGG\_SPEC>3.7).        
\section{Influence of the stellar substructures}
\begin{figure*}
    \centering
    \includegraphics[width = 14cm,clip]{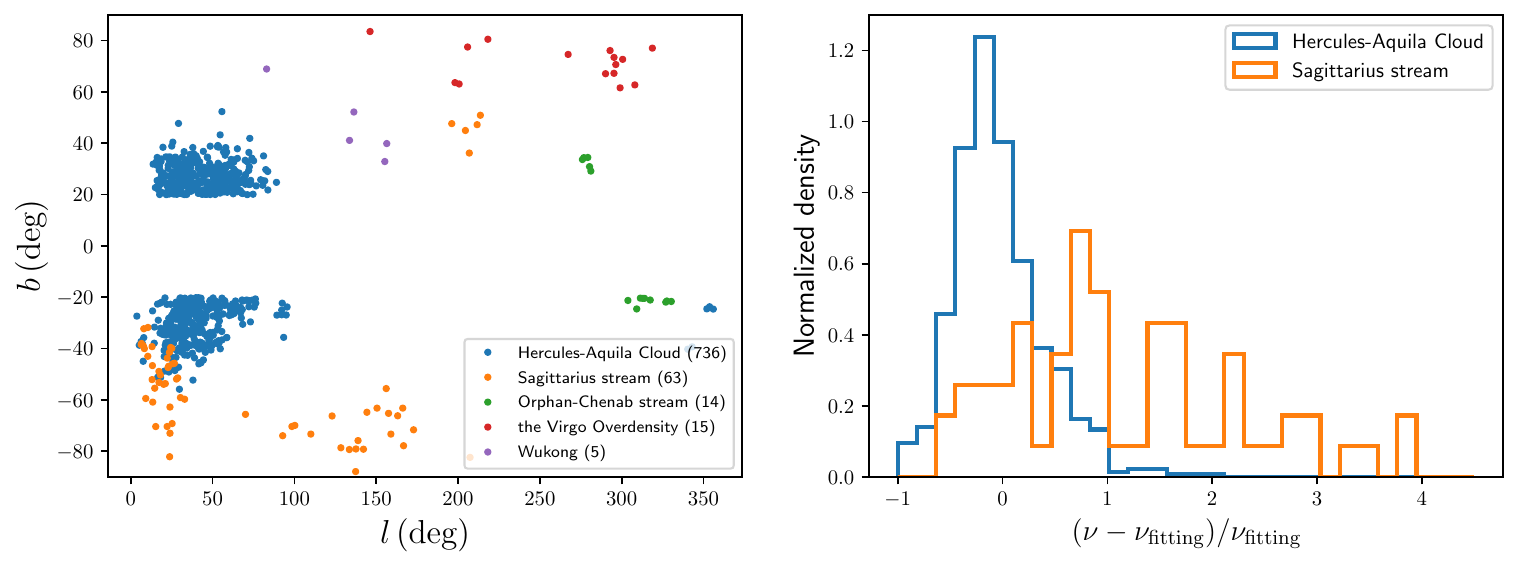}
    \caption{Sky positions in $l$–$b$ space (left panel) are shown for BHB stars that may belong to several halo substructures, along with the normalized distribution of $(\nu - \nu_\mathrm{fitting})/\nu_\mathrm{fitting}$ (right panel) for BHB stars associated with the Hercules-Aquila Cloud and the Sagittarius stream. The numbers in parentheses indicate the number of BHB candidates associated with each substructure.}
    \label{fig:BHBstream}
\end{figure*}

In this appendix, we compared our sample with RR Lyrae members of several known halo substructures identified in \citet{2025ApJ...979..213S}, and selected BHB stars that are similar to these members in both spatial positions (within 1 kpc) and proper motions (within 1 mas yr$^{-1}$). Here we note that this method can only identify a small fraction of BHB stars that may belong to some distinct halo substructures. Figure~\ref{fig:BHBstream} exhibits the distributions of these halo substructures in $l-b$ diagram, along with the normalized distributions of $(\nu-\nu_\mathrm{fitting})/\nu_\mathrm{fitting}$ for BHB stars associated with the two most obvious substructures of the Hercules-Aquila Cloud and the Sagittarius stream. The stellar density $\nu$ does not show a large deviation from $\nu_\mathrm{fitting}$ in the spatial positions corresponding to the Hercules-Aquila Cloud. However, this is not the case in the spatial positions containing BHB stars from the Sagittarius stream, where the mean value of $(\nu-\nu_\mathrm{fitting})/\nu_\mathrm{fitting}$ reaches as high as 1.17. Most BHB stars of the Hercules-Aquila Cloud are located in the inner halo $10<r_\mathrm{gc}<20$ kpc, while the BHB members of the Sagittarius stream are mainly found in the outer halo $r_\mathrm{gc}>20$ kpc. The stellar density is much lower in the outer halo compared to the inner halo. Therefore, if we assume the same increase in stellar density caused by halo substructures, their impact on the derived density profile and shape is likely to be more pronounced in the outer halo. This interpretation is consistent with the residual map shown in Figure~\ref{fig:median}, where the most prominent difference is typically seen in the outer halo.
\end{appendix}
\end{document}